\documentclass[openany]{nature}
\usepackage[T1]{fontenc}
\usepackage[utf8]{inputenc}

\usepackage{overpic}
\usepackage[T1]{fontenc}
\usepackage[left]{lineno}
\usepackage{graphicx}
\usepackage{pdflscape}
\usepackage{hyperref}
\usepackage{threeparttable}
\usepackage{xcolor}
\usepackage{ulem}
\usepackage{cancel}
\usepackage[figuresleft]{rotating}
\usepackage{caption}
\usepackage{subfigure}
\usepackage{longtable}
\usepackage{academicons}
\usepackage{tablefootnote}
\usepackage{threeparttablex}
\usepackage{amsmath, amssymb}
\usepackage{siunitx}
\usepackage{arydshln} 
\def \arcsec {\hbox{$^{\prime\prime}$}}
\usepackage{soul}
\definecolor{orcidlogocol}{HTML}{A6CE39}

\usepackage{subfiles}
\usepackage{hyperref}
\def\UrlAlphabet{%
      \do\a\do\b\do\c\do\d\do\e\do\f\do\g\do\h\do\i\do\j%
      \do\k\do\l\do\m\do\n\do\o\do\p\do\q\do\r\do\s\do\t%
      \do\u\do\v\do\w\do\x\do\y\do\z\do\A\do\B\do\C\do\D%
      \do\E\do\F\do\G\do\H\do\I\do\J\do\K\do\L\do\M\do\N%
      \do\O\do\P\do\Q\do\R\do\S\do\T\do\U\do\V\do\W\do\X%
      \do\Y\do\Z}
\def\UrlDigits{\do\1\do\2\do\3\do\4\do\5\do\6\do\7\do\8\do\9\do\0}

\makeatletter
\g@addto@macro{\UrlBreaks}{\UrlOrds}
\g@addto@macro{\UrlBreaks}{\UrlAlphabet}
\g@addto@macro{\UrlBreaks}{\UrlDigits}

\let\saved@includegraphics\includegraphics
\AtBeginDocument{\let\includegraphics\saved@includegraphics}
\makeatother

\usepackage{booktabs}
\usepackage{multirow}
\usepackage{comment}
\newcommand{\micron}{$\mu$m}

\makeatletter

\renewenvironment*{figure}{\@float{figure}}{\end@float}

\def\@fnsymbol#1{\ensuremath{\ifcase#1\or \dagger\or \ddagger\or
 \mathsection\or \mathparagraph\or \|\or **\or \dagger\dagger
 \or \ddagger\ddagger \else\@ctrerr\fi}}
\makeatother

\definecolor{dkblue}{RGB}{54, 86, 169}
\newcommand{\REV}[1] {{{{#1}}}}
\usepackage{hyperref}
\usepackage{graphicx}
\usepackage{longtable}
\usepackage{supertabular}
\usepackage{float}

\title{A lanthanide-rich kilonova in the aftermath of a long gamma-ray burst}

\author{
Yu-Han Yang$^{1}$\thanks{yuhan.yang@roma2.infn.it}, 
Eleonora Troja$^{1,2}$\thanks{eleonora.troja@uniroma2.it},
Brendan O'Connor$^{3, 4, 5}$,
Chris L. Fryer$^{6,7,8,9,10}$
Myungshin Im$^{11}$,
Joe Durbak$^{4,5}$, 
Gregory S. H. Paek$^{11}$, 
Roberto Ricci$^{12,13}$, 
Cl\'ecio R. Bom$^{14,15}$,
James H. Gillanders$^{1}$,
Alberto J. Castro-Tirado$^{16,17}$,
Zong-Kai Peng$^{18,19}$, 
Simone Dichiara$^{20}$,
Geoffrey Ryan$^{21}$,
Hendrik van Eerten$^{22}$,
Zi-Gao Dai$^{23}$,
Seo-Won Chang$^{11}$, 
Hyeonho Choi$^{11}$, 
Kishalay De$^{24}$,
Youdong Hu$^{16}$, 
Charles D. Kilpatrick$^{25}$,
Alexander Kutyrev$^{4,5}$,  
Mankeun Jeong$^{11}$, 
Chung-Uk Lee$^{26}$, 
Martin Makler$^{27,14}$, 
Felipe Navarete$^{28}$,
Ignacio P\'erez-Garc\'ia$^{16}$
}

\begin{document}

\maketitle

\begin{affiliations}
    \scriptsize
    \item Department of Physics, University of Rome ``Tor Vergata'', via della Ricerca Scientifica 1, I-00133 Rome, Italy
    \item INAF - Istituto Nazionale di Astrofisica, 00133 Rome, Italy
    \item Department of Physics, The George Washington University, 725 21st Street NW, Washington, DC 20052, USA
    \item Department of Astronomy, University of Maryland, College Park, MD 20742-4111, USA
    \item Astrophysics Science Division, NASA Goddard Space Flight Center, 8800 Greenbelt Rd, Greenbelt, MD 20771, USA    
    \item Computer, Computational, and Statistical Sciences Division, Los Alamos National Laboratory, Los Alamos, NM87545, USA
    \item Center for Theoretical Astrophysics, Los Alamos National Laboratory, Los Alamos, NM87545, USA 
    \item The University of Arizona, Tucson, AZ85721, USA 
    \item Department of Physics and Astronomy, The University of New Mexico, Albuquerque, NM87131, USA 
    \item The George Washington University, Washington, DC20052, USA    
    \item SNU Astronomy Research Center, Astronomy Program, Dept. of Physics \& Astronomy, Seoul National University, Seoul 08826, Republic of Korea
    \item Istituto Nazionale di Ricerca Metrologica, I-10135 Torino, Italy
    \item INAF - Istituto di Radioastronomia, via Gobetti 101, I-40129 Bologna, Italy
    \item Centro Brasileiro de Pesquisas F\'isicas, Rua Dr. Xavier Sigaud 150, CEP 22290-180, Rio de Janeiro, RJ, Brazil
    \item Centro Federal de Educa\c{c}\~ao Tecnol\'ogica Celso Suckow da Fonseca, Rodovia M\'ario Covas, lote J2, quadra J, CEP 23810-000, Itagua\'i, RJ, Brazil
    \item Instituto de Astrof\' isica de Andaluc\' ia (IAA-CSIC), Glorieta de la Astronom\' ia s/n, 18080 Granada, Spain
    \item Unidad Asociada al CSIC Departamento de Ingeniería de Sistemas y Autom\' atica, Escuela de Ingenier\' ias Industriales, Universidad de M\'alaga, Arquitecto Francisco Pe\~nalosa, 6, 29071 M\'alaga, Spain
    \item Institute for Frontier in Astronomy and Astrophysics, Beijing Normal University, Beijing 102206, China
    \item Department of Astronomy, Beijing Normal University, Beijing 100875, China 
    \item Department of Astronomy and Astrophysics, The Pennsylvania State University, 525 Davey Lab, University Park, PA 16802, USA
    \item Perimeter Institute for Theoretical Physics, 31 Caroline St. N., Waterloo, ON, N2L 2Y5, Canada
    \item Physics Department, University of Bath, Claverton Down, Bath BA2 7AY, United Kingdom
    \item Department of Astronomy, School of Physical Sciences, University of Science and Technology of China, Hefei 230026, China
    \item Kavli Institute for Astrophysics and Space Research, Massachusetts Institute of Technology, Cambridge, MA, USA    
    \item Center for Interdisciplinary Exploration and Research in Astrophysics (CIERA) and Department of Physics and Astronomy, Northwestern University. Evanston, IL 60208, USA
    \item Korea Astronomy and Space Science Institute, 776 Daedeok-daero, Yuseong-gu, Daejeon 34055, Republic of Korea
    \item International Center for Advanced Studies \& Instituto de Ciencias F\'isicas, ECyT-UNSAM \& CONICET, 1650, Buenos Aires, Argentina
    \item SOAR Telescope/NSF's NOIRLab, Avda Juan Cisternas 1500, 1700000, La Serena, Chile
\end{affiliations}

\clearpage
\begin{abstract} 
Observationally, Kilonovae are astrophysical transients powered by the radioactive decay of nuclei heavier than iron, synthesized in the merger of two compact objects\cite{Eichler89,LiLX1998,
Freiburghaus99, Korobkin2012}. Over the first few days, the kilonova evolution is dominated by a large number of radioactive isotopes contributing to the heating rate\cite{LiLX1998,Barnes2016}. On timescales of weeks to months, its behavior is predicted to differ depending on the ejecta composition and merger remnant\cite{Hotokezaka2020,Zhu2022,Wollaeger2019}. 
Previous work has shown that the kilonova associated with gamma-ray burst 230307A is similar to kilonova AT2017gfo\cite{Levan2023}, and mid-infrared spectra revealed an emission line at 2.15 micrometres that was
attributed to tellurium.
Here we report a multi-wavelength analysis, including publicly available James Webb Space Telescope data\cite{Levan2023} and our own Hubble Space Telescope data, for the same gamma-ray burst. 
We model its evolution up to two months after the burst and show that, at these late times, the recession of the photospheric radius and the rapidly-decaying bolometric luminosity ($L_{\rm bol}\propto t^{-2.7\pm 0.4}$) support the recombination of lanthanide-rich ejecta as they cool. 

\medskip
\end{abstract}

\begin{figure*}
    \centering
    \includegraphics[width=160 mm]{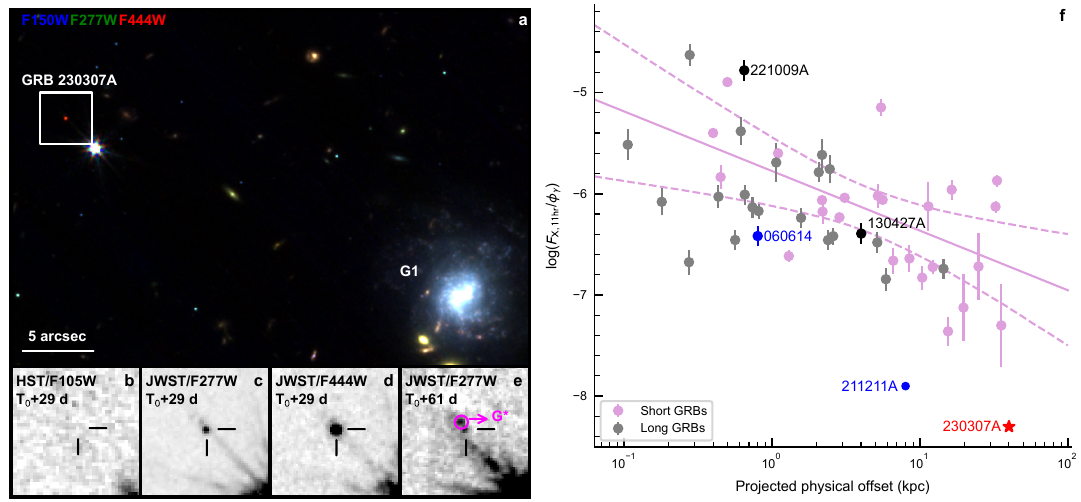}
    \caption{\noindent \textbf{The environment of GRB 230307A. a,} The false-color image combines three filters from \textit{JWST} ($F150W$, $F277W$, and $F444W$). The bright galaxy labeled by G1 is the most likely host galaxy at an offset of $40$ kpc. The white box in (a) shows the zoom on the transient location displayed in panels (\textbf{b, c, d, e}). The transient has a very red color in the near-simultaneous \textit{HST} and \textit{JWST} observations. The high-redshift galaxy G* is marked in the magenta circle in (e). \textbf{f.} Ratio of 0.3--10 keV X-ray flux at 11 hr to the 15--150 keV gamma-ray fluence versus the projected physical offset from the GRB host galaxy. 
    The purple and gray data points represent short and long GRBs, respectively. The purple solid line and dashed lines indicate the best-fit model and 95\% c.l. for short GRBs. The bright long GRBs 221009A and 130427A are shown in black circles. 
    Hybrid long GRBs 060614 and 211211A are shown in blue circles. GRB 230307A is marked by a red star, lying at the bottom of the distribution. Error bars are $1 \sigma$.
}

\label{fig:host}
\end{figure*}

An extremely bright burst, dubbed GRB~230307A, triggered the Gamma-ray Burst Monitor aboard NASA's \textit{Fermi} mission at 15:44:06.67 on March 7th, 2023 (hereafter $T_0$). 
Observationally, GRB~230307A stands out from the general population of long GRBs for three properties: a record-setting gamma-ray fluence\cite{Sun2023} ($\approx$\,$3 \times 10^{-3}$ erg cm$^{-2}$; $10-1,000$ keV), a weak X-ray counterpart (Figure \ref{fig:host}f), and a strong blue-to-red color evolution. 
 
Early observations at optical and near-infrared (nIR) wavelengths identify a weak counterpart, whose brightness ($H\approx 20.2$ AB mag at $T_0$+1.2 d) matches the extrapolation of the X-ray spectrum. The spectral energy distribution (SED) of the GRB counterpart thus does not show evidence for absorption by gas and dust along the sightline (Methods). 
After $T_0$+4 d 
the X-ray and optical emission decay quickly, with temporal power-law indices $\alpha _X = 1.71\pm 0.10$ and $\alpha_O = 2.64 _{-0.26}^{+0.16}$, respectively. Instead, the nIR emission persists for several days after the explosion ($K\simeq$22 AB mag at $T_0$+7 d), and then rapidly declines (Extended Data Figure~\ref{fig:lc_3pl}). Late-time ($\simeq T_0$+29~d) observations with the \textit{Hubble Space Telescope} (\textit{HST}) and \textit{JWST}\cite{Levan2023} show that the peak of the nIR emission shifts from $\approx$22,000 \AA\ at $T_0$+7 d to $\gtrsim$44,000 \AA\ at $T_0$+29 d. At this time ($\simeq T_0$+29~d), the continuum is adequately described by the superposition of a power-law spectrum with spectral index $\beta_{OX} \approx 0.6$ and a blackbody spectrum with temperature $T\approx 638$ K  (observer frame; Methods).

The key ingredient to interpret these observations is the GRB distance scale. Unfortunately, in the case of GRB 230307A, no direct redshift measurement is available.
Our analysis of the photometric dataset provides evidence for a redshift $z\lesssim$3.3 (at the 95\% confidence level, c.l.; Methods). This leaves a range of possible distance scales that is still too broad.
An alternative route to estimate the GRB's distance is to identify its host galaxy using probabilistic arguments\cite{Bloom2002}. In the case of GRB 230307A, this methodology leads to several possible host galaxies: 1) a distant ($z\gtrsim$3.9) star-forming galaxy 
(G$^{*}$ in Figure \ref{fig:host}e); 2) a local origin in the Magellanic Clouds; 
3) a nearby ($z\sim$0.0647, corresponding to 291 Mpc\cite{2019ApJ...882...34F}) face-on spiral galaxy 
(G1 in Figure \ref{fig:host}a). Each of these three possibilities leads to extreme properties for GRB 230307A (Methods).

\begin{figure}
\centering
\subfigure{
\includegraphics[width=160 mm]{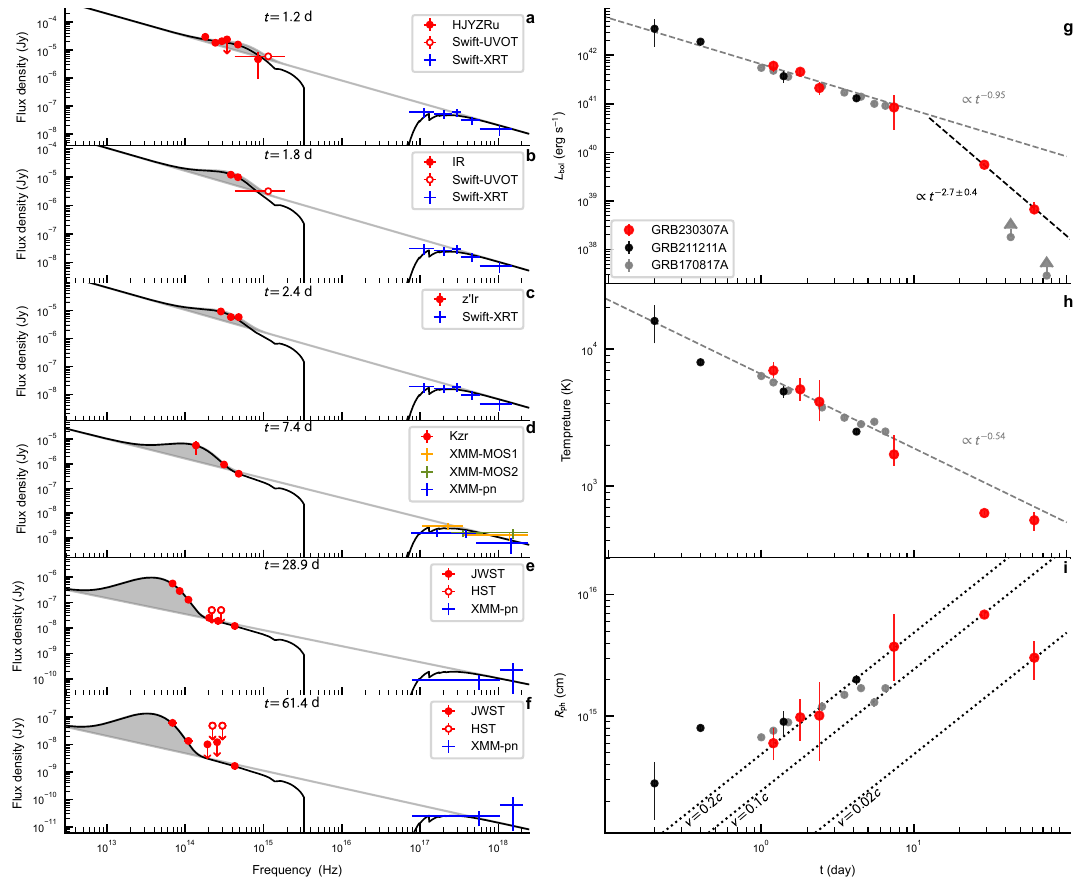}
}
    \caption{\noindent\textbf{Spectral energy distribution of the GRB counterpart.} 
    \textbf{a--f,} The observed spectra at
    different epochs (from top to bottom: 1.2 d, 1.8 d, 2.4 d, 7.4 d, 29 d, 61 d after the GRB)
    are fit with an empirical power-law plus blackbody model (black solid thick lines). 
    The gray shaded areas show the unabsorbed blackbody components, and the gray thin solid lines show the power-law components. Different symbols represent different telescopes or detectors, while the optical data points correspond from left to right according to the label.  \textbf{g-i,} The bolometric luminosity, effective temperature, and photospheric radius of the thermal emission.
    GRB 211211A (black circles)\cite{Troja2022} and GRB 170817A (gray circles)\cite{Waxman2019,Kasliwal2022} are displayed for comparison. 
    The early bolometric luminosity and temperature ($\sim 1-7 $ d) conform to simple power-law decay with slopes $-0.95$ and $-0.54$, respectively\cite{2018MNRAS.481.3423W}. The bolometric luminosity at late times decays with a steeper slope $-2.7\pm 0.4$. The dotted lines in \textbf{i} indicate $R\sim vt/(1+z)$.
    Error bars and upper limits are 1$\sigma$ and 3$\sigma$ c.l., respectively. 
}
\label{fig:sed}
\end{figure}

Further insights can be gleaned from the SED of the GRB counterpart, modeled with a power-law plus a blackbody component (Methods). The results show that a thermal component exists in all spectra acquired $>$\,$T_0$+1 d (Figure \ref{fig:sed}a-f). 
Until $\approx T_0$+10 d, this component exhibits a trend of decreasing temperature and increasing radius, which is consistent with an expanding fireball. By assuming homologous expansion and imposing that the velocity $v\sim (1+z)R / t$ cannot exceed the speed of light, we obtain $z<$0.43 and rule out a high-redshift origin for GRB 230307A. Additionally, at the putative distance of 291 Mpc, the temperature and radius of the thermal component match the evolution of the kilonova counterparts of GRBs 170817A\cite{Waxman2019,Kasliwal2022} and 211211A\cite{Troja2022} (Figure \ref{fig:sed}h,i). This evidence provides additional support to the association between GRB 230307A and G1, and points to a new case of a kilonova following a long-duration GRB \cite{Gehrels2006,Yang060614,Jin060614,Troja2022,2022Natur.612..223R,2022Natur.612..232Y}. 

\begin{figure}
    \centering
    \includegraphics{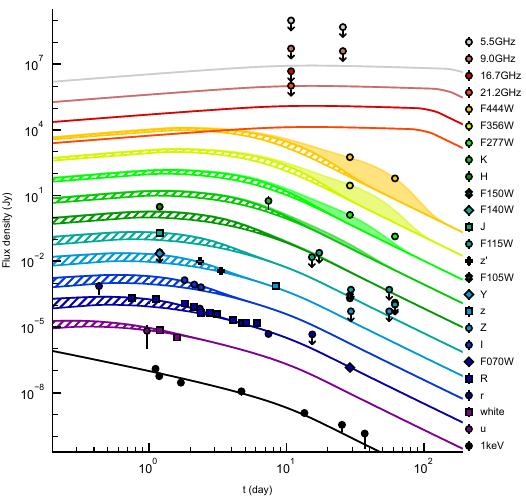}
    \caption{{\bf The multi-wavelength counterpart of GRB 230307A. }  
    The best-fit model to the multi-wavelength lightcurves. Except for X-ray and \textit{u/white}-band, the \textit{n}$^{\textrm{th}}$ lightcurve, when viewed from the bottom upwards, is vertically shifted by a factor of $10^n$. 
    The corresponding observed energy/filter/frequency is shown on the right side. 
    Due to the minor difference in effective wavelengths of some filters, we merged these observations into a single lightcurve.
    These lightcurves were fit by the emission contributed by a GRB afterglow (solid lines) and two-component kilonova (hatched and shaded areas). Error bars and upper limits are 1$\sigma$ c.l. and 3$\sigma$ c.l., respectively. 
       } 
    \label{fig:mwo}
\end{figure}

Based on this classification, we adopt an afterglow plus kilonova model to describe the multi-wavelength counterpart. The afterglow component describes the non-thermal emission arising from the relativistic ejecta and their interaction with the ambient medium\cite{2020ApJ...896..166R}.  The kilonova component instead accounts for the thermal emission arising from the sub-relativistic radioactive ejecta\cite{LiLX1998}. Multiple afterglow solutions are possible (Methods). 
However, regardless of the details of the explosion, the inclusion of a kilonova component represents a significant improvement in all cases ($\Delta {\rm BIC} >140$). For our fiducial afterglow model (Figure \ref{fig:mwo}), we find strong evidence in favor of two kilonova components  over the single kilonova component 
(\REV{$\Delta {\rm BIC} =19$}). The former component is produced by fast moving ejecta ($v \sim 0.2 c$) with mass $M \sim 0.03 M_\odot$ and opacity $\kappa \lesssim 3 {~\rm cm}^2 {\rm ~g}^{-1}$ (3$\sigma$ c.l.). This component mostly contributes to the optical and nIR emission over the first few days, then quickly fades away. The latter component is produced by slightly more massive ($M \sim 0.05 M_\odot$), slower ($v \sim 0.03 c$)  ejecta with a significantly higher opacity ($\kappa \gtrsim 13 {\rm ~cm}^2 {\rm ~g}^{-1}$, 3$\sigma$ c.l.). This component becomes visible after $\sim T_0$+10 d and dominates the late-time emission. Its inclusion is mostly driven by the mid-infrared detections and their steep spectral profile ($\beta_{\rm IR} \approx 3.2$), and relies on the assumption that the contribution of emission lines remains subdominant.

After estimating the contribution of the underlying non-thermal continuum, 
we derive the kilonova properties (Figure \ref{fig:sed}).
At approximately $T_0$+7 d, the effective temperature of the thermal component drops below 2,000 K (Figure \ref{fig:sed}h), and the photospheric layer exhibits a tendency to recede into the inner regions (Figure \ref{fig:sed}i). The velocity distribution as a function of mass can affect the evolution of the photosphere, but would produce a more gradual transition. A similar trend is instead observed in some Type II supernovae (SNe) during their hydrogen recombination phase\cite{Valenti2016}. In the case of a kilonova, the drop in effective temperature changes the ionization states of lanthanides and actinides, transitioning from singly ionized to neutral states, at the critical temperature of around 2,500 K\cite{Barnes2021}. With a lower number of free electrons, the number of infrared bound-bound lines decreases considerably\cite{Frey2013,Fontes2020,Fontes2023}. This causes a drop in the optical depth (Extended Data Figure~\ref{fig:opacity}), accelerating the recession of the photosphere. The outer layers instead enter into an optically thin phase. 
This complex evolution is not accounted for by simple constant-opacity kilonova models and may explain why two kilonova components provide a better description of the dataset.

The kilonova luminosity is seen to rapidly decrease as $L_{\rm bol} \propto t^{-2.7\pm 0.4}$ (Figure \ref{fig:sed}g), ranging from $\sim 6\times 10^{39}$ erg s$^{-1}$ at 29 d to $\sim 7 \times 10^{38}$ erg s$^{-1}$ at 61 d. A rapid decay of the luminosity was identified in the late-time observations of the kilonova AT2017gfo\cite{Kasliwal2022, Waxman2019}, and interpreted as the possible signature of short-lived heavy isotopes dominating the heating rate and thus the observed emission. 
However, in the case of AT2017gfo, the weak signal and limited coverage of the \textit{Spitzer} data were not sufficient to characterize the spectral shape, and only placed lower limits to the true bolometric luminosity. 
In the case of GRB 230307A, 
the sensitivity and multi-color coverage of \textit{JWST} and \textit{HST} observations allow for better sampling.
The data show that the ejecta is only partially optically thin and its late-time nIR luminosity is still dominated by photospheric emission of the inner layers. The evolution of the photosphere is consistent with adiabatic expansion and does not require a drop in the heating rate to explain the change in luminosity. Although this implies that no specific element can be identified based on temporal evolution, the fast decay of the luminosity can still inform us on the properties of this kilonova.

Predictions of the late-time evolution of a kilonova span a wide range of behaviors, depending on nuclear inputs and ejecta properties (e.g. total mass, total energy, velocity distribution, ejecta composition). 
A common expectation is that, if translead nuclei such as $^{254}$Cf are produced in the explosion, their decay products would deposit energy into the ejecta and 
cause the kilonova luminosity to flatten over time\cite{ZhuWollaeger,Holmbeck2023}. A hot central engine (e.g. magnetar, pulsar or fall-back accretion) powering the lightcurve can also alter the late-time emission\cite{Wollaeger2019}.

\begin{figure}
    \centering
    \includegraphics{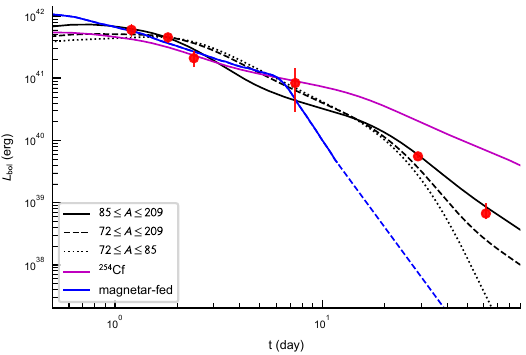}
    \caption{\textbf{Comparison of the bolometric lightcurve to different models.}  
    The black lines are calculated using the model from ref.\cite{Hotokezaka2020} with the solar r-process abundance pattern of different atomic mass ranges of $85\leq A \leq209$ (black solid line), $72\leq A \leq209$ (black dashed line) and $72\leq A \leq85$ (black dotted line). The ejecta parameters are $M_{\rm ej} = 0.07M_\odot$, $v_{\rm ej} = 0.1c$, $v_{\rm ej,max} = 0.4c$, and $\beta_{v} = 1.5$. The opacity is $\kappa _1 = 0.6$ cm$^2$ g $^{-1}$ and $\kappa _2 = 20$ cm$^2$ g $^{-1}$ with the velocity threshold $0.15c$.
    \REV{The purple solid line illustrates the effects of efficient energy deposition by the spontaneous fission of $^{254}$Cf, which is calculated by the same method and parameters as black lines except for $M_{\rm ej}=0.05 M_\odot$, and directly incorporating effective heating rate with $^{254}$Cf from ref. \cite{ZhuWollaeger}.} 
    The blue line with dashed extrapolation illustrates the generic evolution of a magnetar-fed kilonova. \REV{We show the same model used to describe AT2017gfo with $M_{\rm ej} = 0.001M_\odot$, $v_{\rm ej,max} /2=0.45c$, characteristic spindown luminosity $L_0=2\times 10^{44}$ erg s$^{-1}$,  gravitational wave dominated spindown timescale $t_{\rm gw}=495$ s and magnetar lifetime $t_{\rm cut}=23$ d \cite{Wollaeger2019}.} } \label{fig:Lbol}
\end{figure}

By comparing the bolometric lightcurve with different models (Figure \ref{fig:Lbol}),  we find that the efficient energy deposition of a long-lasting magnetar\cite{Wollaeger2019} or actinide fission fragments\cite{ZhuWollaeger,Holmbeck2023} is not consistent with the red color and rapid decay of the bolometric light curve. 
A radioactive-powered kilonova containing r-process elements beyond the first peak (\REV{atomic mass number} $A \gtrsim 85$) display a better agreement with the data. This is because lighter elements have shorter lifetimes and cannot provide sufficient radioactive power at these late epochs, resulting in a dimmer and cooler kilonova. The bolometric lightcurve, coupled with the observed evolution of the photospheric radius and the inferred high opacity, points to lanthanide production in the merger ejecta, and confirms kilonovae are a cosmic site of heavy r-process elements.

\clearpage

\clearpage

\clearpage

\clearpage

\clearpage
\section*{Methods}
\subsection{Spectral Energy Distribution\\}
We used \texttt{XSPEC}\cite{Arnaud1996}
to jointly fit the near-infrared, optical and X-ray SEDs at 1.2, 1.8, 2.4, 7.4, 28.9 and 61.4 d (Figure \ref{fig:sed} and Extended Data Table \ref{tab:empfit}). 
The observed optical data were converted to spectral files using the \texttt{ftflx2xsp} and \texttt{uvot2pha} tools within HEASOFT v6.31. 
Necessary data were extrapolated based on observations at nearby times and fit results of empirical lightcurve modelling (Extended Data Figure \ref{fig:lc_3pl} and Supplementary Methods).
The Galactic contribution was modeled using the model \texttt{phabs} for X-ray photoelectric absorption with a fixed parameter $N_H = 1.26 \times 10^{21}$ cm$^{-2}$ (ref.\cite{2013MNRAS.431..394W}), and the model \texttt{redden} for optical dust reddening\cite{Schlafly2011} with fixed parameters $E(B - V) = 0.0758$ mag. 
Each SED was fit using a power-law (PL) model and a blackbody (BB) plus PL model. 

To constrain the presence of absorption systems at the GRB site, we include two additional components (\texttt{zphabs} and \texttt{zdust}), and varied the GRB distance scale up to a redshift $z<0.5$. 
The fit with a PL function yields tight constraints ($N_{H,z}< 2.5 \times 10^{21}$ cm$^{-2}$ and $E(B-V)_z<0.03$ mag at 1.2 d, 3$\sigma$ c.l.), which disfavor
models of GRB afterglows interacting with a dusty ambient medium\cite{Waxman2022}. 
However, the results show positive evidence for the inclusion of a BB component, starting as early as $T_0$+1.2 d. The best-fit parameters are listed in Extended Data Table \ref{tab:empfit}.
The addition of a thermal component loosens the constraints on the rest-frame reddening ($E(B-V)_z<0.3$ mag), but its 
rapid onset remains inconsistent with the typical timescales of a thermal dust echo.

\subsection{Constraints to the GRB Distance Scale\\}
The GRB distance scale can leave a detectable imprint on its afterglow SED\cite{Kruhler11}, therefore we include in our spectral fits three components that are sensitive to the GRB redshift: \texttt{zphabs} and \texttt{zdust} to model the absorption within the GRB host galaxy, 
and \texttt{zigm} to describe the effects of the intervening intergalactic medium. 
By selecting 100 redshift values uniformly distributed between 0 and 10, we mapped the variation of the test statistics ($\chi ^2$) as a function of redshift. Analyzing this sample statistic allowed us to derive an upper limit 
of $z<3.3$ at the 95\% c.l. ($z<4.1$ at the 99.9\% c.l.). 
This result is mostly driven by the afterglow detection in the \textit{Swift}/UVOT filters $u$ and $white$, which disfavor a high-redshift origin for GRB 230307A. 

An even more stringent constraint can be placed using the properties of the BB component. 
From the fit, we derive its effective temperature $T_{\rm eff}$ and total observed flux, which can then be used to derive the bolometric luminosity $L_{\rm bol}$ and, assuming isotropy\cite{2023Natur.614..436S}, the radius $R_{\rm ph}$ of the emitting surface for a certain redshift. 
By imposing that the expansion velocity $v \sim (1+z)R_{\rm ph}/t$ cannot exceed the speed of light, we obtain $z < 0.23$ and $z < 0.43$ from the SED at 1.2 d and 28.9 d, respectively.

\subsection{Host Galaxy Association\\}
Here we explore the GRB environment using deep multi-color imaging with \textit{HST} (observed on April 6, 2023) and \textit{JWST} (observed on May 8, 2023). 
A faint galaxy, referred to as G* (Figure \ref{fig:host}e), is located $0.24 \pm 0.01\arcsec$ from the explosion site and lies at $z$\,$\gtrsim$\,$3.9$ 
based on the identification of a H$\alpha$ emission line in its spectrum \cite{2023GCN.33580....1L}. 
For this galaxy, we derive magnitudes $F070W>28.3$ mag, $F115W>28.6$ mag, $F150W>28.8$ mag, $F277W=27.80\pm0.15$ mag, and $F444W=28.15\pm0.20$ mag. 
Using the galaxy number counts from the \textit{JWST} Prime Extragalactic Areas for Reionization and Lensing Science (PEARLS) project\cite{Windhorst2023}, we estimate a probability of chance coincidence\cite{Bloom2002} $P_{cc}\gtrsim0.04$ using the $F150W$ limit, and a $P_{cc}\approx0.03$ using the $F277W$ brightness. 

A high-redshift origin would fit well within the classification of GRB 230307A as a long burst from a young stellar population. 
However, a redshift of $z \gtrsim 3.9$ is in tension with the limits from optical measurements ($z\lesssim$3.3). Moreover, it would imply an unprecedented gamma-ray energy release ($\sim 10^{56}$ erg) followed by the onset of an extremely luminous and blue ($M_g \approx -25.7$ at 7~d) transient, which has never been observed before.  These considerations lead us to disfavor the association between the GRB and the distant galaxy G$^{*}$.

The next most likely associations, in terms of posterior probability, are the Magellanic Clouds, which lie $\approx$8 degrees away from the GRB.  By cross-correlating the catalog of \textit{Swift}
bursts with the 10 brightest galaxies of the Local Group, we derive $P_{cc}\sim 0.05$ from the number of GRBs located within an 8-degree radius of any of those galaxies. 
However, a distance scale of only 50 kpc would drive GRB 230307A to 
the low-luminosity extreme of high-energy transients, consistent with the population of giant flares (GFs) from soft gamma-ray repeaters, rather than GRBs. 
These GFs are characterized by a quasi-thermal spectrum\cite{2002JASS...19....1C}, which is not observed in GRB 230307A\cite{dichiara23}. Unlike GRB 200425A, a candidate extra-galactic GF\cite{2020ApJ...899..106Y}, the low-energy photon index of GRB 230307A is $\approx$\,$-1$\cite{dichiara23,WangY2023}, consistent with the non-thermal shape of GRBs. 
Based on the properties of its high-energy emission, a local origin for GRB~230307A is likewise disfavored.

We perform a further exploration of the field surrounding GRB 230307A using our deep \textit{HST} imaging in order to determine whether there is any other probable host. We computed the offset and \textit{HST}/$F140W$ photometry for all extended sources in the field. In Extended Data Figure \ref{fig:galaxy}a, we display the probability of chance coincidence for a sample of nearby galaxies versus their distance from the GRB's localization. While there exist a handful of additional galaxies within 10\arcsec\ of the GRB position, there are no galaxies with similarly low probabilities. 
Due to their faintness, each of these nearby galaxies has $P_{cc}>0.25$ making them unlikely hosts. 
However, we identify a bright galaxy, hereafter G1 (Figure \ref{fig:host}a) at an offset of $30\arcsec$ from the GRB localization. 
This galaxy has an infrared brightness of $F140W\sim17.6$ AB mag, which yields a $P_{cc}=0.13$. 

From the spectrum of G1 (Extended Data Figure \ref{fig:galaxy}b), we derive a redshift of $z=0.0647\pm 0.0003$ for this galaxy based on H$\alpha$, H$\beta$, [OIII], [NII], and [SII] emission lines. 
Therefore, we derive a nearby distance of 291 Mpc for G1, assuming a $\Lambda$CDM cosmology with a Hubble constant of $H_0 = 69.8$ km Mpc$^{-1}$ s$^{-1}$, $\Omega _M =0.315$ and $\Omega _\Lambda =0.685$\cite{2019ApJ...882...34F, 2020A&A...641A...6P}.
The projected physical distance between the GRB and G1 would be $\approx$40 kpc, among the largest values measured for GRBs\cite{OConnor2022}, yet not unprecedented among short GRBs\cite{Jin070809}. 
This seems consistent with the low X-ray flux to gamma-ray fluence ratio, also observed for GRB 211211A\cite{Troja2022}, and interpreted as an indication of a burst in a low-density environment\cite{OConnor2020}. 
However, even this interpretation is not free from uncertainties. In the case of GRB 230307A, the complex morphology of the gamma-ray lightcurve hardly resembles any of the previous examples of short GRBs with extended emission\cite{Norris2006,dichiara21,dichiara23}. Additionally, the probability of chance alignment between the GRB and the nearby galaxy G1 is $P_{cc}$\,$\approx$13\%, a value generally considered too high for a reliable physical association.

In what follows, we consider G1 the most likely host for GRB 230307A, as, despite the high $P_{cc}$, the study of the afterglow SED points towards a low redshift origin. 
The emission line properties of G1, namely L$_{\textrm{H}\alpha}\approx4\times10^{40}$ erg s$^{-1}$ and $\log([\textrm{NII}]/$H$\alpha)\approx-0.6$, 
provide estimates of the star formation rate (SFR) and metallicity of the galaxy, leading to SFR$\approx0.2$ $M_\odot$ yr$^{-1}$ \cite{Kennicutt1998,Chabrier2003} and 12 + $\log$ O/H\, $\approx$ 8.6\cite{Kobulnicky2004}.

We modeled the SED of G1 (Extended Data Table \ref{tab:galaxymags}) using \texttt{prospector}\cite{Johnson2021} with the same setup as previously outlined in ref.\cite{OConnor2021,OConnor2022,Troja2022}. The data were first corrected for Galactic extinction along the line of sight\cite{Schlafly2011}. Our best fit model is displayed in Extended Data Figure \ref{fig:galaxy}c. We derive a stellar mass of $M_*/M_\odot=(2.4\pm0.9)\times10^{9}$, a star formation rate of $0.20\pm0.03$ $M_\odot$ yr$^{-1}$, an intrinsic dust component with extinction of $A_V=0.20\pm0.02$ mag, 
a metallicity $Z/Z_\odot = 0.04^{+0.02}_{-0.01}$, and a mass-weighted stellar age of $2.8^{+2.2}_{-1.5}$ Gyr. The specific star formation rate sSFR $\approx$ 0.3 Gyr$^{-1}$ is low for a long GRB host galaxy\cite{Palmerio2019}. In fact, the host galaxy properties as a whole (a low-mass galaxy, a low star formation rate, and an old stellar population) point towards a host galaxy that is entering quiescence\cite{Whitaker2012}. This is quite similar to the host galaxy of GRB 211211A\cite{Troja2022}. These similarities in the inferred host properties highlight a growing population of long-duration GRBs produced by the merger of two compact objects that may occur in similar galaxy types.

\subsection{Prompt Emission\\}
Additional constraints on the GRB nature and its distance scale can be placed by a study of its prompt gamma-ray emission, shown in Extended Data Figure \ref{fig:prompt}. 
By converting the total fluence $3\times 10^{-3}$ erg cm$^{-2}$ s$^{-1}$ (10--1,000 keV)\cite{Sun2023} 
to an isotropic energy ($E_{{\rm iso}, \gamma}$) and peak energy $1255$ keV into rest-frame $E_p(1+z)$, the red dashed/solid line in Extended Data Figure \ref{fig:prompt}d illustrates GRB 230307A at different redshifts on the Amati-relation diagram\cite{2002A&A...390...81A}. 
For a wide range of typical GRB redshifts ($0.25<z<1.7$) GRB 230307A fits within the $1\sigma$ c.l. region of the standard distribution for long GRBs.
For redshifts $0.01<z<0.06$, it falls within the distribution of short GRBs (95\% c.l.), whereas for even lower distances its isotropic energy approaches the SGR region, 
orders of magnitude lower than the weakest GRB\cite{2020ApJ...899..106Y}.
Extended Data Figure \ref{fig:prompt}d shows that, if associated with the most likely host galaxy G$^*$ at $z\sim$3.9, GRB 230307A would not only be the most energetic explosion ever observed, with an energy release an order of magnitude higher than that of GRB 221009A\cite{Oconnor2023}, 
but would also deviate significantly from the general population of long GRBs. This provides us with additional evidence against a high-redshift origin. 
More plausible values ($E_{\gamma,{\rm iso}} \approx 3\times 10^{52}$ erg) are found assuming the distance of G1 at $z\sim0.0647$.  However, the GRB lies at the intersection between the two populations of bursts and the Amati diagram does not reduce the uncertainty in its classification.

\subsection{Possible Progenitors for GRB 230307A\\}
The traditional classification of GRBs divides them into either long or short GRBs, based on a threshold of 2 s\cite{Kouveliotou93}, which are generally associated with the collapse of massive stars and compact binary mergers (including NS-BH, NS-NS), respectively. 
The merger of a white dwarf (WD) with a NS was also proposed to explain long GRBs without an associated SN\cite{2022Natur.612..232Y,Becerra2023}.

\textit{Massive star collapse:} 
GRB 230307A is a burst of long duration, commonly associated with the collapse of a massive star. For a broad range of redshifts ($0.25<z<1.7$), the prompt emission of GRB~230307A fits within the Amati relation for long GRBs (Extended Data Figure \ref{fig:prompt}d),  typically followed by a SN. 
Using the prototypical SN1998bw\cite{2011AJ....141..163C} 
and the faint SN2022xiw\cite{Srinivasaragavan2023}, 
as templates and comparing them to the deep \textit{HST} and \textit{JWST} observations (Extended Data Figure \ref{fig:snz}),  we rule out the possibility of a SN out to $z>6.8$ and $z>3.3$, respectively.  
Similarly, we eliminate fast-evolving SNe, such as SN2005E\cite{Perets2010} and SN2010X\cite{Kasliwal2010}, at $z<1$. 
Given the constraint $z <0.43$ based on the evolution of the BB component, we find no plausible range of redshift and extinction values that could accommodate a massive star progenitor.

\textit{White Dwarf/Neutron star merger:} 
The merger of a WD--NS binary system can give rise to a GRB, provided the WD is sufficiently massive (e.g. GRB 211211A\cite{2022Natur.612..232Y,Zhong2023}). In such scenarios, the timescale of the GRB may extend beyond 2 s\cite{1999ApJ...520..650F,Kaltenborn2023}. The presence of neutron-rich matter\cite{2022Natur.612..232Y} or materials undergoing radioactive decay\cite{Zhong2023} in the ejecta can contribute to additional optical excess\cite{2022MNRAS.510.3758B}, alongside the standard GRB afterglow. 
A NS-WD merger is a plausible origin for GRB 230307A, and can explain many of its unusual properties, from the long duration to its environment. However, existing models for the associated optical transient \cite{Kaltenborn2023,2022MNRAS.510.3758B} do not predict the rapid reddening observed in GRB 211211A and GRB 230307A, and match more closely the evolution of faint Type Iax SNe rather than kilonovae. Based on this fact, we tend to favor a compact binary merger as progenitor for GRB 230307A.

\textit{Compact binary merger (NS-NS/BH):} 
The merger of two compact objects, comprising at least one NS, 
is known to produce a GRB and a short-lived red thermal transient, a kilonova\cite{Jin070809}.
Although it is challenging to conceive that the duration of GRBs originating from these mergers can extend to tens of seconds\cite{LuW2023}, 
this progenitor system best explains
the properties of the GRB counterpart, such as its very red color and rapid evolution, and its environment.

\subsection{Multi-wavelength Afterglow Modelling\\} 

The non-thermal afterglow radiation that follows GRBs is best described as synchrotron emission from a population of shock-accelerated electrons. 
Multiple mechanisms contribute to shaping its evolution, the dominant being an external forward shock (FS) driven by the interaction of the GRB jet with the ambient medium. 
A reverse shock (RS) traveling backward into the ejecta or long-lasting activity of the central engine may also contribute to the afterglow emission at early times\cite{Gao2013}. 

Due to the delayed localization of GRB~230307A, the available dataset does not allow us to unambiguously identify the origin of its early non-thermal emission. 
Therefore, we consider two 
possible options in our modeling: a) FS radiation is the only dominant component at all times, and b) a RS and/or central engine activity contributes at early times, with the FS dominating later on ($>1$ d). In the former case, we include all observational data (Dataset 1) in our fit, in the latter case we consider the \textit{TESS} and ATCA detections as upper limits (Dataset 2) to the FS radiation. 

Motivated by the results of the SED analysis, we fit these datasets with three models: \textit{i}) a simple FS model, \textit{ii}) a FS plus a kilonova (KN) component, and \textit{iii}) a FS plus a two-component KN. 
In order to model the FS emission, we utilized the Python package \texttt{afterglowpy}\cite{2020ApJ...896..166R}. The free parameters are the isotropic-equivalent kinetic energy $E_0$, the circumburst density $n_0$, the fraction of burst kinetic energy in magnetic fields $\varepsilon_B$ and in electrons $\varepsilon_e$, the power-law slope $p$ of the electron energy distribution, the opening angle of the jet's core $\theta_c$, and the electron participation fraction $\xi_N$. We applied a Gaussian structure for the GRB jet $E(\theta)=E_0 \exp(-\theta^2/2\theta_c^2)$ for $\theta\leq\theta_w$, where $\theta_w=4\,\theta_c$ is the truncation angle. 
Due to the extreme brightness of GRB 230307A's prompt emission, we consider only an on-axis viewing angle $\theta_v\approx 0$ rad.

Regarding the kilonova component, we employed the isotropic model from ref.~\cite{Metzger2019} in the Python package \texttt{gwemlightcurves}. 
The model assumes a gray opacity and describes the spectrum with a simple blackbody function. 
The free parameters in this model are the ejecta mass $M_\textrm{ej}$, its minimum velocity $v_\textrm{ej}$, velocity index $\beta_v$, opacity $\kappa$, and electron fraction $Y_\textrm{e}$.

We fit the multi-wavelength observations based on the nested sampling algorithm implemented in the Python package \texttt{pymultinest}\cite{2014A&A...564A.125B}. 
Considering that our data come from a variety of different telescopes, we included an additional 5\% systematic uncertainty for all observations. 
The best-fit parameters, resulting $\chi^2$ and Bayesian information criterion (BIC) for each model and dataset are shown in Extended Data Tables \ref{tab:fitstat}, \ref{tab:fskn_fit} and Extended Data Figure \ref{fig:FSKNfit}. 
The FS model resulted in all cases in a poor description of the data ($\chi^2$/dof$>$4), 
and the addition of a KN component substantially improves the fit ($\chi^2$/dof$\approx$2).

The fit to Dataset 1 is tightly constrained by the early optical lightcurve, which determines the peak flux and frequency of the FS component. 
It requires a very early jet-break ($\sim 0.3$ d) to accommodate the rapid decay of the broadband emission, thus stretching the physical parameters to unusual values. 
For these reasons, our fiducial model is based on the fit to Dataset 2 (Figure \ref{fig:mwo}). 
A two-component kilonova provides a better description
($\chi^2$/dof$\approx$1.5) of the optical and infrared excess, capturing the full blue-to-red evolution of the transient, and the bright mid-infrared emission at late times.

\subsection{Opacity Evolution\\}
In transient explosions, the photosphere moves inward in mass coordinates. 
But because the ejecta is moving rapidly outwards, the radius of the photosphere likewise increases. In Type II supernovae, the opacity in the ejected hydrogen envelope is dominated by electron scattering. 
When these electrons recombine, the drop in opacity accelerates the recession of the photosphere in mass coordinates.  The subsequent deceleration of the radial expansion is a key feature in understanding the plateau phase of Type IIP supernovae\cite{Valenti2016}.  In kilonovae, the opacity is dominated by bound-bound line transitions in lanthanides.  Assuming local thermodynamic equilibrium, Extended Data Figure~\ref{fig:opacity} shows the different opacities for neodymium (Nd) as it evolves from singly-ionized Nd, to neutral Nd.  The number of lines in the 1-5\,\micron\ range drops considerably as it recombines, and the overall opacity will also decrease by roughly an order of magnitude.  This sudden decrease in opacity will accelerate the inward motion of the photosphere in mass coordinates, causing the radius of the photosphere to stop increasing, or even to recede.  Nonthermal effects can alter these opacities in Extended Data Figure~\ref{fig:opacity}, but the basic trends with decreasing temperature are likely to hold.

\section*{Data Availability}
\textit{Swift}/XRT products are available from the online GRB repository (\url{https://www.swift.ac.uk/xrt\_products}).
\textit{Swift}/UVOT data are available from Swift Data Access (\url{https://www.swift.ac.uk/archive}).
X-shooter data are available from ESO Science Archive Facility (\url{https://archive.eso.org}).
\textit{HST} and \textit{JWST} data are available from Mikulski Archive for Space Telescopes (\url{https://mast.stsci.edu}).
\textit{Chandra} data are available from Chandra Data Archive (\url{https://cda.harvard.edu/chaser}).
The \textit{TESS} lightcurve is available from TessTransients archive
(\url{https://tess.mit.edu/public/tesstransients}).
Gemini data are available from Gemini Observatory Archive (\url{https://archive.gemini.edu}).
\textit{XMM-Newton} data are available from \textit{XMM-Newton} Science Archive (\url{https://www.cosmos.esa.int/web/xmm-newton/xsa}).
\REV{
\textit{Fermi}/GBM data are available from Fermi Science Support Center (FSSC) FTP archive  \url{https://heasarc.gsfc.nasa.gov/FTP/fermi/data/gbm}.
Source data in Figures 1f, 2 and 3 are provided. All the processed data are available upon request to the corresponding authors.}

\section*{Code Availability}
Results can be reproduced using standard free analysis packages.
Methods are fully described. Codes used to produce figures can be made available upon request.

\begin{addendum}
\item
This work was supported by the European Research Council through the Consolidator grant BHianca (grant agreement ID 101002761) and, in part, by the National Science Foundation (under award number 2108950). 
This work was in part carried out at the Aspen Center for Physics, which is supported by National Science Foundation grant PHY-2210452. 
The development of afterglow models used in this work was partially supported by the European Union Horizon 2020 Programme under the AHEAD2020 project (grant agreement number 871158).
B.O. acknowledges useful discussions with Justin Pierel and Ori Fox regarding \textit{JWST} analysis. 
M.I., G.S.-H.P., S.-W.C., H.C., and M.J. acknowledge support from the National Research Foundation of Korea (NRF) grants, No. 2020R1A2C3011091, and No. 2021M3F7A1084525, funded by the Korea government (MSIT).
C.R.B. acknowledges the financial support from CNPq (316072/2021-4) and from FAPERJ (grants 201.456/2022 and 210.330/2022) and the FINEP contract 01.22.0505.00 (ref.1891/22). C.R.B. made use of HPC Sci-Mind servers machines developed and supported by the CBPF AI LAB team. This research has made use of the KMTNet system operated by the Korea Astronomy and Space Science Institute (KASI) at three host sites of CTIO in Chile, SAAO in South Africa, and SSO in Australia. Data transfer from the host site to KASI and SNU was supported by the Korea Research Environment Open NETwork (KREONET).
\REV{A.J.C.-T. acknowledges funding of the Spanish Ministry project PID2020-118491GB-I00/AEI/10.13039/501100011033 project.}
The national facility capability for SkyMapper has been funded through ARC LIEF grant LE130100104 from the Australian Research Council, awarded to the University of Sydney, the Australian National University, Swinburne University of Technology, the University of Queensland, the University of Western Australia, the University of Melbourne, Curtin University of Technology, Monash University and the Australian Astronomical Observatory. SkyMapper is owned and operated by The Australian National University's Research School of Astronomy and Astrophysics. The survey data were processed and provided by the SkyMapper Team at ANU. The SkyMapper node of the All-Sky Virtual Observatory (ASVO) is hosted at the National Computational Infrastructure (NCI). Development and support of the SkyMapper node of the ASVO has been funded in part by Astronomy Australia Limited (AAL) and the Australian Government through the Commonwealth's Education Investment Fund (EIF) and National Collaborative Research Infrastructure Strategy (NCRIS), particularly the National eResearch Collaboration Tools and Resources (NeCTAR) and the Australian National Data Service Projects (ANDS).

\item[Author Contributions] 
Y.-H.Y led the analysis of the prompt emission, SEDs, and multi-wavelength lightcurves. 
E. T. initiated the project, coordinated the observations and their interpretation. 
B.O. led the study of the host galaxy.
C.R.B., A.J.C.-T., Y.H., C.D.K., M.M., F.N., I.P.-G., and J.H.G acquired and reduced the data of the SOAR telescope. 
R.R. led the analysis of the radio data. 
B.O. reduced the \textit{JWST} data. 
E.T. and B.O. acquired and reduced the \textit{HST} data. 
E.T., B.O. and Y.-H.Y acquired and reduced the \textit{XMM-Newton} data. 
E.T. and Y.-H.Y reduced the \textit{Swift} data. 
Y.-H.Y reduced the TESS data. 
M.I, G.S.-H.P., M.J., S.-W.C., H.C. and C.-U.L. acquired and reduced the data of the KMTNet and RASA36 telescopes. 
E. T., J.D., K. D., A. K. and S.B.C. acquired and reduced the data of the PRIME telescope. 
B.O., S.D. and J.H.G. acquired and reduced the data of the Gemini telescope. 
J.H.G. and E.T. reduced the data of the X-shooter telescope. 
Y.-H.Y, G.R., H.v.E., Z.-G.D. contributed to afterglow modelling and their physical interpretation. 
Z.-K.P. contributed to possible progenitors. 
C.L.F. contributed to the interpretation of the data. 
Y.-H.Y, E.T., B.O. and C.L.F. wrote the manuscript, with contributions from all authors. 

\item[Competing Interests] The authors declare that they have no competing financial interests. Supplementary Information is available for this paper.

\item[Correspondence] Correspondence and requests for materials should be addressed to Y.-H. Yang \\
(yuhan.yang@roma2.infn.it) and E. Troja (eleonora.troja@uniroma2.it).

\end{addendum}
\clearpage
\setcounter{figure}{0}
\captionsetup[figure]{name={\bf Extended Data Figure}}

\clearpage

\begin{figure}
\centering
\includegraphics{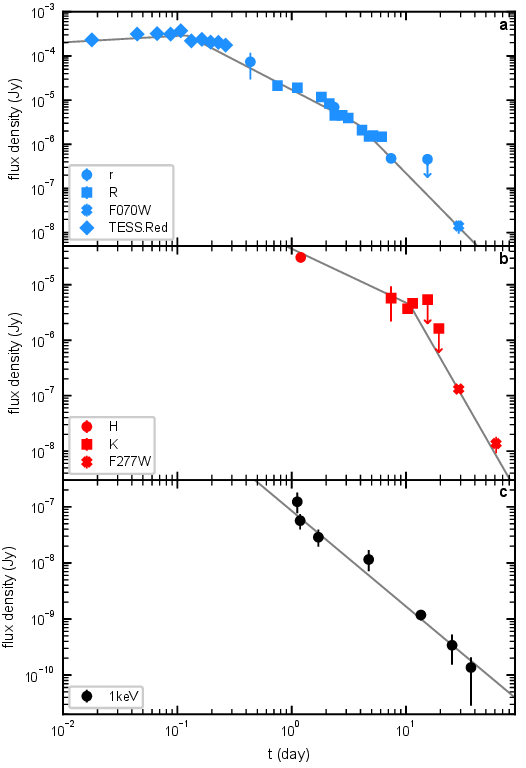}
\caption{\noindent\textbf{ Empirical model for the nIR, optical, and X-ray lightcurves.}  The lightcurves are modeled using PL segments, $F_\nu \propto t^{-\alpha} \nu^{-0.8}$. The gray lines represent the best-fit models.
Different symbols indicate observations with different filters. Error bars and upper limits are 1$\sigma$ c.l. and 3$\sigma$ c.l., respectively. 
} 
\label{fig:lc_3pl}
\end{figure}

\begin{figure}
\centering
\includegraphics[width=160 mm]{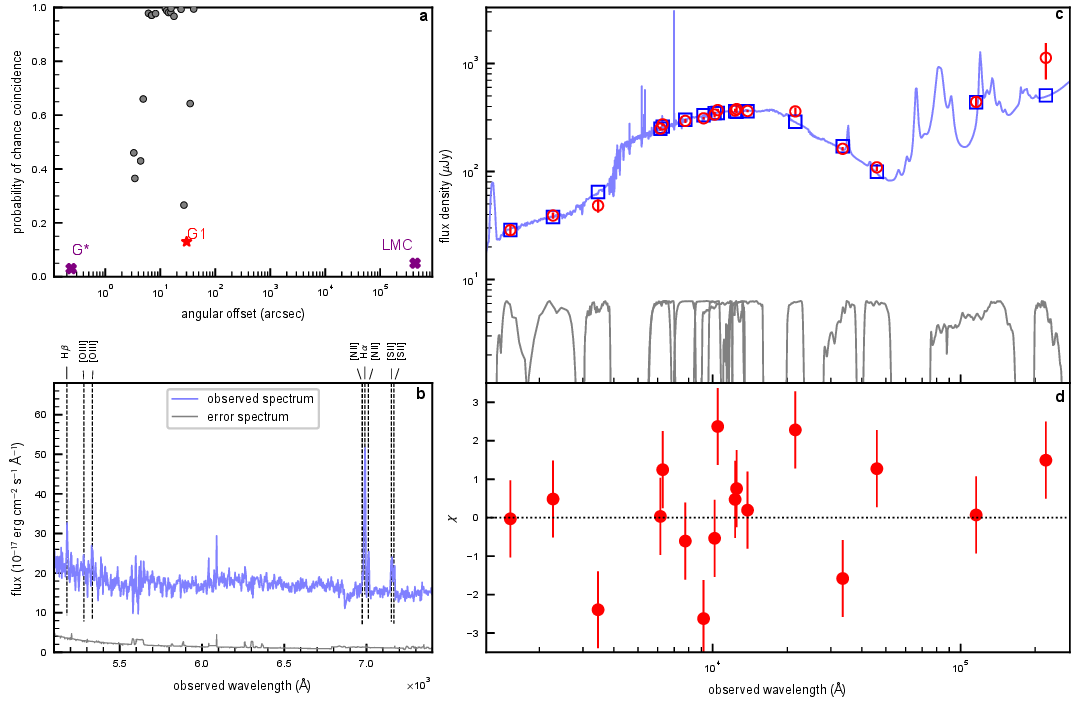}
\caption{\noindent \textbf{Properties of the potential host galaxies.}
\textbf{a,} Probability of chance coincidence for galaxies in the field of GRB 230307A. Likely unrelated galaxies are displayed as gray circles. The candidate host galaxies G*, LMC (purple crosses) and G1 (red star) are highlighted.  
\textbf{b,} Optical spectrum of the bright galaxy G1. The observed spectrum is shown in blue and the error spectrum in black. Line identifications are made at $z=0.0647\pm0.0003$. The spectrum is smoothed with a Savitzky-Golay filter of two pixels for display purposes.  
\textbf{c,d,} Spectral energy distribution of the bright galaxy G1. The model SED (blue line) and model photometry (blue squares) derived using \texttt{Prospector} are compared to the observed photometry (red circles). Filter bandpasses are shown at the bottom of panel \textbf{c} in gray. 
Fit residuals are shown in \textbf{d}. Error bars represent $1\sigma$ uncertainties.
}
\label{fig:galaxy}
\end{figure}

\begin{figure}
\centering
\subfigure{\includegraphics[width=160 mm]{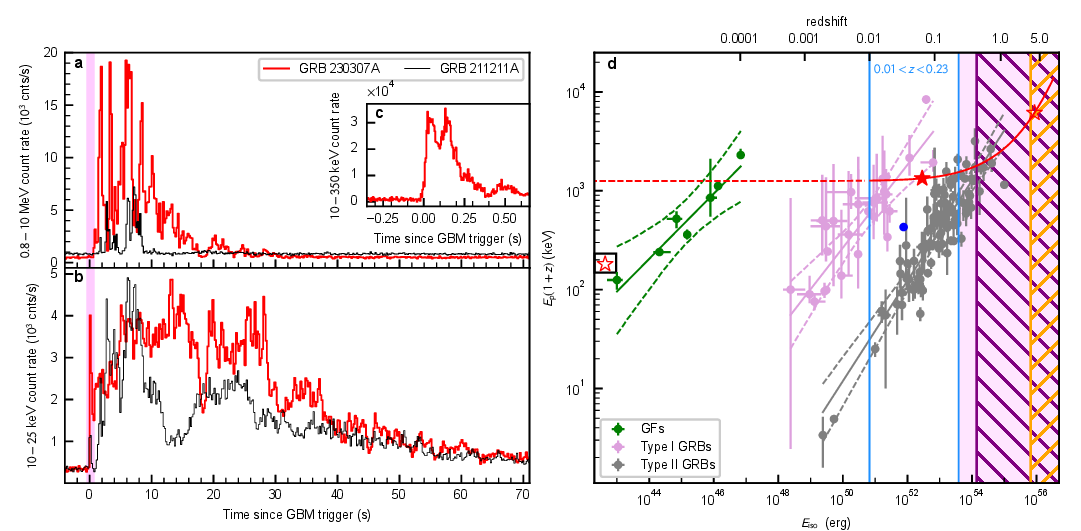}}
\caption{\noindent\textbf{Prompt emission properties of GRB 230307A.} \textbf{a,b,} Gamma-ray lightcurves of GRB 230307A (red) and GRB 211211A (dark) from \textit{Fermi}/GBM in the energy range of $10-25$ keV and $0.8-10$ MeV with 0.2 s binsize. The purple shaded area roughly represents the time range of the initial pulse of the lightcurve, as depicted in the zoomed-in panel (\textbf{c}) with 5 ms binsize in the energy range 10--350 keV. \textbf{d,} The Amati-relation diagram. The plum/gray/green circles represent Type I (short) GRBs/Type II (long) GRBs/magnetar giant flares, and the corresponding color solid line and the area between dashed lines are the best-fit model and 95\% c.l., respectively. GRB 230307A (whole burst) shifts following the red line when located at different redshifts. The red stars represent it at the three most probable host galaxies (G1, LMC and G*), while the GF is only reasonable when we treat the initial pulse as the main burst (zoom-in panel \textbf{c}). Hybrid GRB 211211A is shown in the blue circle. 
The purple shaded ($z > 0.23$)/hatched ($z > 0.43$) area is ruled out by the expansion velocity of the photosphere radius at $T_0+$1.2 d/28.9 d being limited to less than the speed of light.
The orange hatched area is ruled out by the SED ($z \lesssim 3.3$). The red dashed line indicates the redshift where it departs from the 95\% c.l. for the distribution of Type I GRBs. Error bars represent $1\sigma$ uncertainties.
\label{fig:prompt}
}
\end{figure}

\begin{figure}
\centering
\includegraphics[width=160 mm]{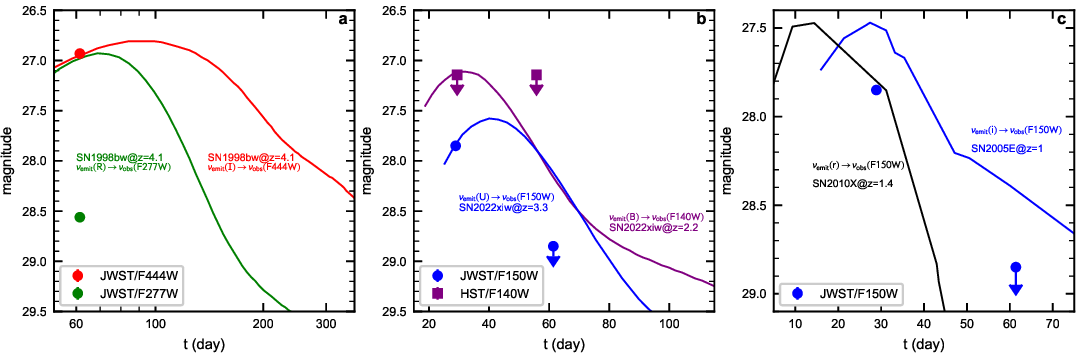}
\caption{\noindent\textbf{ Comparison of \textit{JWST} and \textit{HST} observations and supernova lightcurves at different redshifts.} 
The lightcurves of SN1998bw/GRB 980425\cite{2011AJ....141..163C} (\textbf{a}), SN2022xiw/GRB 221009A\cite{Srinivasaragavan2023} (\textbf{b}), SN2005E\cite{Perets2010} and SN2010X\cite{Kasliwal2010} (\textbf{c}) are employed as references for typical GRB-SNe, bright GRB-SNe and fast-evolving SNe. Error bars and upper limits are 1$\sigma$ c.l. and 3$\sigma$ c.l., respectively.}  

\label{fig:snz}
\end{figure}

\clearpage
\begin{figure}
\includegraphics[width=160 mm]{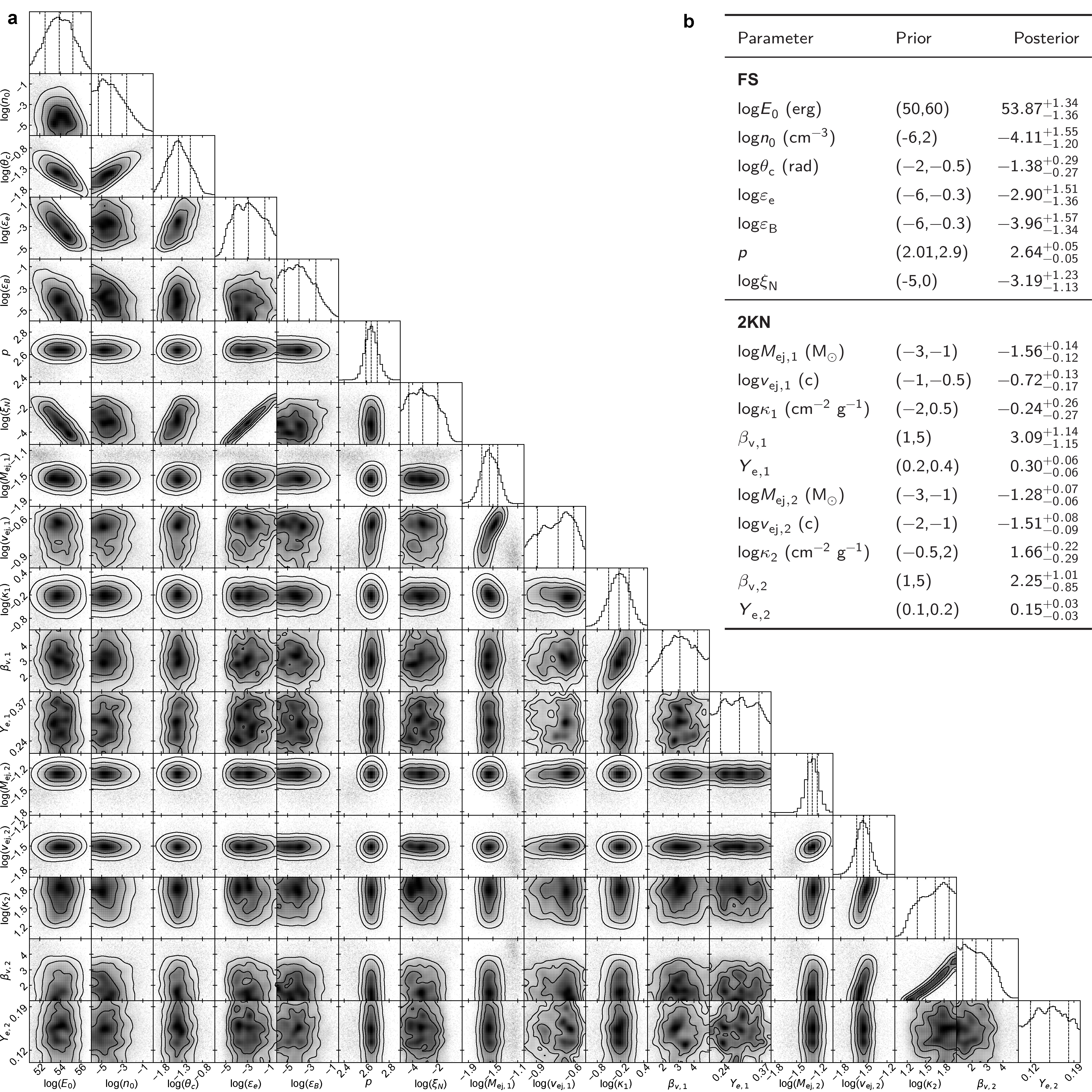}
\caption{{\bf Results for a forward shock plus two-component kilonova model for Dataset 2. a. } Posterior probability distributions of parameters. {\bf b.} The prior bounds and posterior medians for parameters. The values corresponding to the two kilonova components are denoted by the subscript 1 or 2. \REV{Uniform priors are employed for all parameters except for electron index $p$, which is a truncated-Gaussian prior ($2.46\pm0.20$) derived by introducing the spectral index $\beta _X =0.73\pm0.10$\cite{Mereghetti2023} into the closure relations\cite{ZhangMeszaros2004}.} Error bars represent $1\sigma$ uncertainties.
} 
\label{fig:FSKNfit}
\end{figure}

\clearpage
\begin{figure}
\begin{center}
\includegraphics{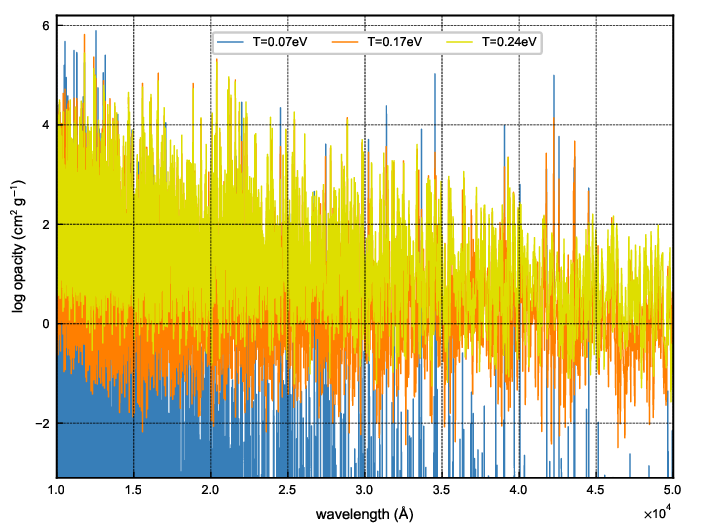}
\end{center}
\caption{\noindent \textbf{Neodymium opacities in the $1-5$ \micron \, range at 3 temperatures:  0.24 eV, 0.17 eV and 0.07 eV.}  In local thermodynamic equilibrium, these correspond to ionization fractions of 1.0 ($T=0.24$ eV), 0.886 ($T=0.17$ eV) and $10^{-6}$ ($T=0.07$ eV).  The material begins to recombine between 0.24 and 0.17 eV (2,000--2,500 K).  As it recombines, the number of bound-bound lines in the 1--5 \micron \, range decreases significantly, causing a drop in the opacity.}
\label{fig:opacity}
\end{figure}

\setcounter{figure}{0}
\captionsetup[figure]{name={\bf Extended Data Table}}
\clearpage
\begin{figure}
\centering
\includegraphics{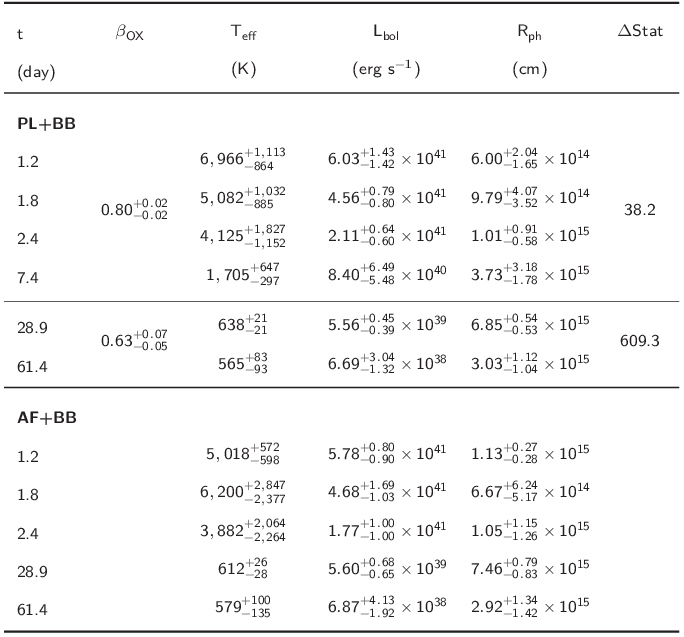}
\caption{\textbf{Infrared/optical/X-ray joint spectral fit results for power-law or afterglow plus blackbody model.} Errors represent the $1\sigma$ uncertainties. The bolometric luminosity and photosphere radius are calculated based on the assumed luminosity distance $D_L = 291$ Mpc. The fit statistics between the power-law (PL) model and PL plus blackbody (BB) model, $\Delta {\rm Stat} = {\rm Stat}_{\rm PL} - {\rm Stat}_{\rm PL+BB}$, represents the improvement in spectral fit by an additional BB component.
The afterglow components are generated based on our fiducial model.}
\label{tab:empfit}
\end{figure}

\clearpage
\begin{figure}
\centering
\includegraphics{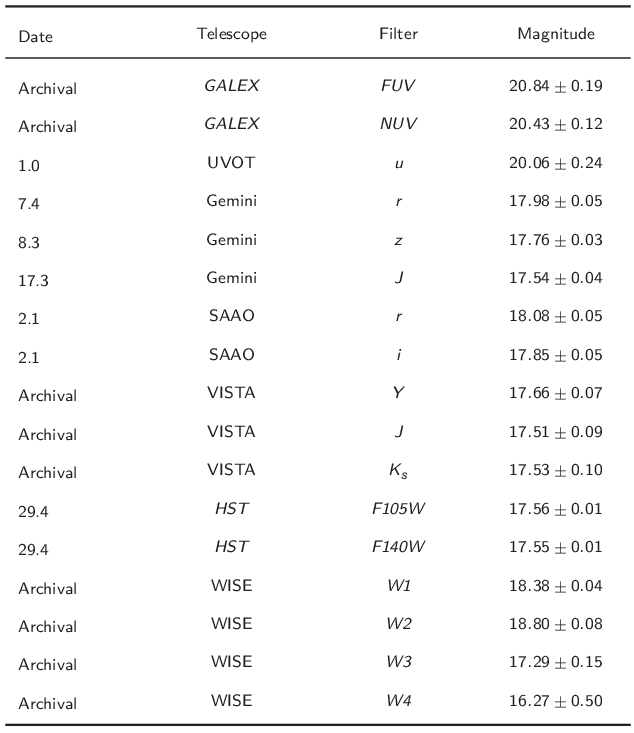}
\caption{\textbf{Photometry of the bright galaxy G1.} Magnitudes are reported in the AB system and are not corrected for Galactic extinction. Errors represent the $1\sigma$ uncertainties. }
\label{tab:galaxymags}
\end{figure}

\clearpage
\begin{figure}
\centering
\includegraphics{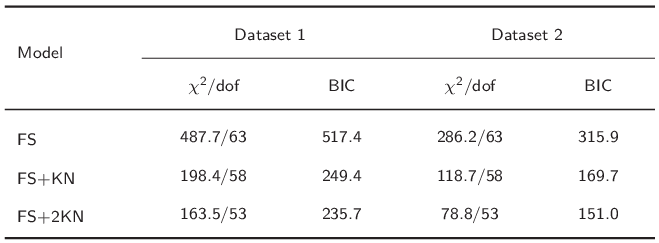}
\caption{
\textbf{Summary of model fitting results.} 
The fit statistics for the three different models analyzed in this work: a forward shock (FS) model, a forward shock and a kilonova (FS+KN), and a forward shock with a two-component kilonova (FS+2KN). All available observational data are included in Dataset 1. Compared to Dataset 1, the \textit{TESS} and ATCA detections are treated as upper limits in Dataset 2. \REV{The best-fit $\chi^2$ divided by the degree of freedom (dof) and Bayesian information criterion (BIC) obtained from all model fittings are compared.}}
\label{tab:fitstat}
\end{figure}

\clearpage
\begin{figure}
\centering
\includegraphics{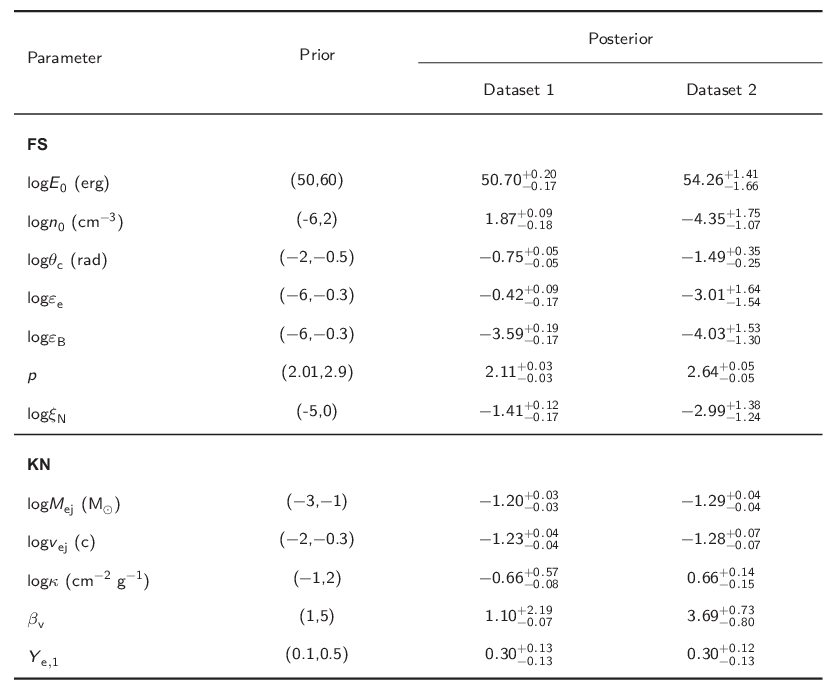}
\caption{\textbf{\REV{Model parameters, prior bounds and posterior medians} from modeling of Dataset 1 and Dataset 2 with a forward shock (FS) plus a kilonova model (KN).} \REV{A truncated-Gaussian prior distribution with mean 2.46 and standard deviation 0.20 is used for electron index $p$, while uniform priors are employed for other parameters.} Errors represent the $1\sigma$ uncertainties.}
\label{tab:fskn_fit}
\end{figure}
\clearpage
\setcounter{page}{1}
\section*{Supplementary Information}
\subsection{X-ray Observations\\}
GRB 230307A was observed with the \textit{Neil Gehrels Swift Observatory} X-ray Telescope (XRT), \textit{XMM-Newton} (PIs: E. Troja and B. O'Connor) and \textit{Chandra} (PI: Fong). We downloaded the XRT spectra from the \textit{Swift}/XRT GRB repository. 
\textit{XMM-Newton} data was reduced by the Science Analysis System (\texttt{SAS}) v21.0. \textit{Chandra} data was reduced by \texttt{CIAO} v4.15.
Details for \textit{XMM-Newton} and \textit{Chandra} observations are listed in Supplementary Table \ref{tab:xraydetail}. This GRB was detected in \textit{Chandra} observation, the first \textit{XMM-Newton} observation (by all detectors), and the second \textit{XMM-Newton} observation (only by detector pn).
We modeled the spectra with an absorbed power-law model with fixed Galactic hydrogen column density $N_{\rm H}=1.26\times 10^{21}$ cm$^{-2}$ (ref. \cite{2013MNRAS.431..394W}).
Assuming the X-ray photon index is constant, we conducted a joint analysis of the XRT and \textit{XMM-Newton} (the detected detectors) spectra using \texttt{XSPEC} v12.13. We derived the photon index of $\Gamma_X=1.80\pm 0.17$. 
We retrieved the XRT lightcurve from the \textit{Swift}/XRT GRB Lightcurve Repository. The XRT lightcurve was further rebinned to a minimum of 8 counts per bin. 
The unabsorbed flux (0.3--10 keV) and flux density (1 keV) are shown in Supplementary Table \ref{tab:xraysum}.

\subsection{Optical and Infrared Observations\\}  
GRB 230307A was observed with \textit{Transiting Exoplanet Survey Satellite} (\textit{TESS}), the PRime-focus Infrared Microlensing Experiment (PRIME) telescope, the \textit{Swift} 
UltraViolet and Optical Telescope (UVOT), the Korea Microlensing Telescope Network (KMTNet), the 8.1m Gemini South Telescope (PIs: B. O'Connor and S. Dichiara; GS-2023A-DD-104, GS-2023A-DD-106), the 4.1m Southern Astrophysical Research (SOAR; PI: C. R. Bom) Telescope, the Very Large Telescope (VLT) X-shooter instrument, the \textit{HST} (PI: E. Troja; ObsID: 17298) and \textit{JWST} (PI: A. Levan; PIDs: 4434 and 4445). The results are presented in Supplementary Table~\ref{tab:photometry - UVOIR}, and details of the observations and data processing are reported below.

\textit{TESS:} \textit{TESS} observed GRB\,230307A during the prompt emission phase and the afterglow phase immediately after the burst\cite{Fausnaugh2023}. The data are presented in ref. \cite{Fausnaugh2023}. We rebinned its early afterglow lightcurve with a minimum of 20,000 counts. 

\textit{KMTNet} and  \textit{RASA36:}
We observed GRB 230327A with the the RASA36 telescope in Chile\cite{SomangNet} of the Gravitational-wave Em Counterpart Korean Observatory\cite{GECKO} 
and three 1.6m telescopes (SAAO, SSO, and CTIO) of KMTNet\cite{KMTNET}. 
Observations were carried out between March 8 and March 14, 2023 (from $T_0$+0.4 d to $T_0$+6.1 d ) in the $r$, $R$ and $I$ filters. 
Photometric measurements were obtained via aperture photometry using \texttt{photutils}\cite{photutils} and calibrated against the SkyMapper catalog \cite{SkyMapper2007, SkyMapper_DR2_2019}. 
The target was observed at an average airmass of 1.7 with variable seeing of $\approx$0.7" (SAAO), $\approx$0.9-1.0" (SSO, CTIO), and $\approx$3.9" (RASA36).

\textit{Gemini:} 
Imaging observations were carried out in the $rzJ$ filters at 2.4, 7.4, 8.3, 15.3, and 17.3 d after the GRB trigger with seeing $\approx$0.8", except for the observations at $T_0+15.3$ d characterized by slightly worse conditions (seeing $\approx$1.0"). 
Imaging observations were analyzed using \texttt{DRAGONS}\cite{DRAGONS_OG_paper, DRAGONS}, and photometry was performed using \texttt{SExtractor}\cite{SExtractor} and calibrated to nearby reference stars from the SkyMapper catalog\cite{SkyMapper2007, SkyMapper_DR2_2019} and the Two Micron All Sky Survey (2MASS) catalog \cite{Skrutskie2006}.

\textit{X-shooter:}
X-shooter\cite{Vernet2011} obtained a spectrum of GRB\,230307A through Target of Opportunity (ToO) observations (ToO 110.24CF.001; PI: N. Tanvir) on 15 March 2023, at 00:18:41 UTC ($\sim 7.4$ d after the GRB trigger). 
The observations had a total integration time of 4800\,s, and spectra were recorded in all three spectrographs (the UVB, VIS and NIR arms). 
We reduced this observation using standard recipes with the European Southern Observatory (ESO) Reflex pipeline\cite{esoreflex_pipeline_paper}.
We did not detect the transient in either of the UVB or VIS arms, but recorded a weak trace in the NIR arm ($\sim K$-band). We derive an estimate of the $K$-band magnitude at this time by folding the spectrum with the filter response. 
Our estimate is confirmed by the detection of nIR emission in Gemini images at 10 d post-merger \cite{Levan2023}, which, unfortunately, was never publicly released in time to allow for activation of approved programs on kilonovae. 

\textit{SOAR:}
We obtained $i$- and $z$-band imaging with the Goodman High Throughput Spectrograph on the SOAR telescope on 9, 10, 17, and 24 March 2023.
The seeing of these images was 1.2", 1.0", 0.8", and 2.3", respectively. 
Following standard procedures in {\tt photpipe}\cite{2005ApJ...634.1103R}, we reduced these data using the same methods described in ref.\cite{Gordon2023}, including corrections for bias and flat-fielding, astrometric calibration, image stacking using {\tt SWarp}\cite{2002ASPC..281..228B}, point-spread function photometry using {\tt DoPhot}\cite{1993PASP..105.1342S}, and photometric calibration against the SkyMapper catalog.  Finally, we used the stacked $z$-band image from 17 March 2023 as a template to obtain deep limits on transient emission at the site of GRB 230307A from 9 and 10 March 2023.  After using {\tt HOTPANTS} to subtract the template image from the science frames on both dates, we performed forced PSF photometry with {\tt DoPhot} at the afterglow coordinates to estimate limits on the brightness of the counterpart. 

\textit{PRIME:} 
The 1.8m \textit{PRIME}\cite{2023JAI....1250004Y} telescope observed GRB 230307A beginning on March 8, 2023 at 18:31:50 UT until 22:45:13 UT using the near-infrared prime focus camera. Observations were conducted in the $ZYJH$ filters with seeing of 1.7", 1.7", 1.5" and 2.0", respectively. 
The sample up-the-ramp data was reduced into single exposures using a memory-efficient up-the-ramp cosmic ray rejecting algorithm\cite{2005PASP..117...94O}. The ramp-fit exposures were reduced using standard photodiode array processing techniques of flat-fielding and background subtraction, aligned using {\tt Astrometry.net}\cite{Lang2010}, and stacked using {\tt SWarp}\cite{2002ASPC..281..228B}. 
Aperture photometry was calibrated to 2MASS ($J$ and $H$ filters), VISTA ($Y$ filter) and SkyMapper ($Z$ filter). 

\textit{Swift/UVOT:}
The \textit{Swift}/UVOT observed the field of GRB 230307A beginning on 14:57:20.33 and 19:22:25.15 on 8 March 2033 using \textit{u} and \textit{white} filters, respectively. 
Magnitudes were calculated using the \textit{Swift} tool \texttt{uvotdetect} by selecting a circular source region with a 3" radius and nearby source-free background region.

\textit{HST:} 
We imaged the field with the \textit{HST} using the WFC3/IR camera with the $F105W$ and $F140W$ filters. Two epochs of observations were performed on April 6 and May 2, 2023, corresponding to 29.4 and 55.8 d after the GRB trigger. We reduced and analyzed the data using standard \texttt{DrizzlePac} procedures. Photometry was performed using \texttt{SExtractor} with the zeropoints derived from the image headers.

\textit{JWST:}
We analyzed publicly available \textit{JWST} observations obtained on April 5, 2023 and May 8, 2023. Imaging observations were performed with NIRCam, using the $F070W$, $F115W$, $F150W$, $F277W$, $F356W$ and $F444W$ filters in the first epoch and $F115W$, $F150W$, $F277W$, and $F444W$ in the second epoch. 
We retrieved the pipeline processed Level-3 mosaics from the MAST Archive. In the first epoch, a clear source is visible in the NIRCam observations at the position of GRB 230307A (Figure 1).
In the second epoch, the source has faded significantly and is only clearly detected in the redder filters ($F277W$ and $F444W$). 
We further retrieved the calibrated Level-2 data products and performed point-spread-function (PSF) photometry using \texttt{WebbPSF}\cite{Perrin2014} and \texttt{photutils}\cite{photutils}. We applied the zeropoints derived from the keywords in the image headers. We further verified that simple aperture photometry obtained similar results.

\subsection{Radio Observations\\}
The Australia Telescope Compact Array (ATCA) observed GRB 230307A as part of project CX529 on 12 March, 18 March and 2 April 2023. The array operated in continuum mode with the Compact Array Broadband Backend using 2048 channels over 2-GHz bandwidth per IF. The radio sources used as bandpass/primary and phase calibrator were 1934-638 and 0454-810. The phase calibrator was also used for reference pointing checks every hour. Standard procedures in the data reduction software package \textit{MIRIAD}\cite{1995ASPC...77..433S} were used to flag, calibrate the complex gains and image the phase calibrator and target. A robustness parameter value of two was used, and a 6-km baseline antenna was deselected during imaging. The resulting flux densities and 3$\sigma$ upper limits are presented in Supplementary Table \ref{tab:radio}. 
The errors are derived from summing in quadrature the statistical and systematic uncertainty (map rms and 5\% of the total flux as residual of gain calibration, respectively).

\subsection{Empirical Modelling\\}
We describe the temporal evolution of the GRB counterpart using a series of PL segments, $F_\nu \propto t^{-\alpha}$ 
(Extended Data Figure 1).
The optical band (including filters $r$, $R$, $F070W$ and \textit{TESS}/$Red$) is best fit ($\Delta {\rm BIC} = 138$) 
by three PL segments with indices $\alpha _{\rm O,1} = -0.15 _{-0.14}^{+0.13}$, $\alpha _{\rm O,2} = 1.35 _{-0.10}^{+0.05} $, $\alpha _{\rm O,3} = 2.64 _{-0.26}^{+0.16}$, and temporal breaks
$t_{b,1} = 0.12 _{-0.01}^{+0.04}$ d, and $t_{b,2} = 3.79 _{-0.37}^{+1.13}$ d.  
The nIR lightcurves (including filters \textit{H}, \textit{K}\cite{Levan2023} and \textit{F277W})
is best described by two PL segments with $\alpha_{\rm nIR,1} = 0.97_{-0.30}^{+0.14}$, $\alpha_{\rm nIR,2} =3.60_{-0.27}^{+0.42}$ and a break time $t_b=10.72_{-3.70}^{+1.48}$. 
The X-ray lightcurve is instead satisfactorily described by a single PL function with index $\alpha_X = 1.71\pm 0.10$. 

Additionally, we use the measured X-ray flux 
at $T_0+11$~hr ($F_{\rm X, 11hr}\sim 3\times 10^{-12}$ erg cm$^{-2}$ s$^{-1}$, 0.3-10 keV band), and gamma-ray fluence ($\Phi _\gamma \sim 6\times 10^{-4}$ erg cm$^{-2}$, 15--150 keV band, estimated according to ref.\cite{Sun2023}) 
to compare GRB 230307A to the broader GRB population. 
As shown in Figure 1f, we compared the ratio of the X-ray flux at 11 hr to the gamma-ray fluence and the offset from the center of the host galaxy with those of other short GRBs\cite{OConnor2022} and a random sample of long GRBs\cite{2016ApJ...817..144B}.
Similarly to GRB 211211A \cite{Troja2022}, GRB 230307A exhibits a large deviation from the typical location of GRBs on the diagram, and its log$(F_{\rm X, 11hr}/\Phi _\gamma)\approx -8.3$ is even lower than GRB 211211A ($\approx -7.9)$, which potentially indicates a commonality in the nature of these hybrid GRBs. Therefore, the ratio of these two parameters provides a useful diagnostic tool for identifying outliers in the long GRB population, which may be candidate hybrid GRBs. 

\subsection{Kilonova Bolometric Luminosity\\}
\textit{Effective heating from the spontaneous fission of Californium-254:}
If the magnetic fields behind the shock are turbulent, a generic assumption -- made by most decay heating routines used in kilonova models\cite{ZhuWollaeger,Holmbeck2023} -- is that the energy released from charged particle emission from the decay of $^{254}$Cf is deposited in the ejecta. 
Efficient energy deposition from actinides fission would cause a flattening of the late-time bolometric luminosity.
Alternatively, for a different configuration of the magnetic field (e.g. radial), $^{254}$Cf decay might not efficiently deposit its energy into the ejecta. 
\REV{As shown in Figure \ref{fig:Lbol}, $^{254}$Cf yield as high as $1.5 \times 10^{-7} M_\odot$ \cite{ZhuWollaeger} fits the data well in the first 10 d, but to match the weak signal detected by JWST at 61 d, the yield must either be an order of magnitude lower or the energy deposition from $^{254}$Cf must decrease by a factor of 10 due to an increased escape fraction of its decay products. The low $^{254}$Cf abundance suggests low fission barriers near the predicted $N=184$ shell closure along the $^{254}$Cf $\beta$-feeding path \cite{Vassh2019,Zhu2021}.}
\\
\textit{Magnetar-fed kilonova:}
The long duration of GRB 230307A, as well as its soft X-ray plateau, can be interpreted as the spin-down emission from a magnetar remnant\cite{Sun2023}. 
The electron fraction of the ejecta driven by a hot neutrino-heated magnetar wind is generally too high for robust r-process production. If we assume that the NS remnant lifetime is tracked by the duration of the high-energy emission ($>20$ s), then $Y_e\,\gtrsim 0.4$\cite{MTQ2018}. Only light r-process elements, characterized by a low opacity, would be produced by this outflow. The resulting kilonova is blue because of the high temperature and low opacity of the ejecta. As the photosphere recedes within the ejecta rapidly (especially in mass space), this blue component will become apparent. 
Although it might take a few days before observing the effects of the inner engine\cite{Wollaeger2019}, the fact that the kilonova progressively becomes redder up to $T_0$+61 d, and that we do not see a blue component from the uncovering of a long-lived central engine argues against energy deposition from a remnant magnetar.
To be consistent with the observed data, a magnetar central engine would require that its wind mostly flows into the funnels created by the relativistic jets\cite{WangYH2023}, or the heating efficiency is very low from a Poynting-flux-dominated wind\cite{AiSK2022}, so that energy injection is substantially weakened without substantially affecting the composition of the sub-relativistic ejecta.
\REV{Assuming the early emission originates from contributions by a magnetar, the same model used to describe AT2017gfo with ejecta mass $M_{\rm ej}=0.001 M_\odot$ and characteristic spindown luminosity $L_0=2\times 10^{44}$ erg s$^{-1}$ \cite{Wollaeger2019} could also be applicable here. The magnetic field strength is $\sim 3\times 10^{12}$ G with typical neutron star radius $1.2\times 10^6$ cm and initial rotation period $P=1$ ms.
A higher magnetic field strength is required if the remnant lifetime is shorter, or its rotation is slower. When increasing the ejecta mass, the luminosity at $\sim 1$ d decreases due to increased optical depth, however, delayed emission flattens the luminosity decline, i.e., the luminosity at $\sim 7$ d increases \cite{Wollaeger2019}.}

\clearpage
\setcounter{table}{0}
\captionsetup[table]{name={\bf Supplementary Table}}

\clearpage
\begin{table}
\small
\caption{\textbf{Log of X-ray observations.}} 
\label{tab:xraydetail}
\centering
\begin{threeparttable} 
\begin{tabular}{lccccccc}
\toprule
Telescope & ObsID & Date$^a$ & Detector & Exposure time &  Counts & $R_{\rm src}$ \\
  & &  & &(ks)   \\ 
\midrule 
\multirow{6}{*}{\textit{XMM-Newton}} &\multirow{3}{*}{0915391601} &\multirow{3}{*}{2023-03-20}& pn & 36.4$^b$ & 73.2$^c$ (41.0$^d$) & \multirow{3}{*}{12\arcsec} \\
 & && MOS1 & 36.4& 32.4 (14.3) &  \\
 & && MOS2 & 38.7 & 36.5 (14.0)  &  \\
\cmidrule{2-7}
&\multirow{3}{*}{0915391701} &\multirow{3}{*}{2023-04-13}& pn & 85.0 & 132.6 (108.7)& \multirow{3}{*}{12\arcsec} \\
 & && MOS1 & 105.8& 43.8 (42.8) &  \\
 & && MOS2 & 114.2 & 37.7 (38.9)&  \\
 \midrule
\multirow{3}{*}{\textit{Chandra}} &27777 & 2023-03-31& ACIS-S&19.8 &1 (0.04) & \multirow{3}{*}{1\arcsec}\\
 &27778 & 2023-04-01& ACIS-S&16.9 &4 (0.04) & \\
 &27779 & 2023-04-02& ACIS-S&13.6 &0 (0.05) & \\
\bottomrule
\end{tabular}

\clearpage
\begin{tablenotes}
\footnotesize
\item [a] Start date of the observation.
\item [b] The exposure times in good time intervals (GTI), i.e. excluding high-rate flaring particle background.  
\item [c] Total counts in the circle source region with the radius $R_{\rm src}$.
\item [d] Background counts in the source region.
\end{tablenotes}
\end{threeparttable}
\end{table}
\begin{table}
\caption{\textbf{X-ray lightcurve of GRB 230307A.} Errors represent the $1\sigma$ uncertainties.}
\label{tab:xraysum}
\centering
\begin{threeparttable} 
\begin{tabular}{cccc}
\toprule
Time  & Telescope &  Flux$^a$ & Flux density$^b$ \\
(d) &   &(erg cm$^{-2}$ s$^{-1}$) &  (Jy) \\ 
\midrule 
$1.12_{-0.15}^{+0.04}$&\textit{Swift}/XRT & $1.2_{-0.4}^{+0.5}\times 10^{-12}$ &$1.2_{-0.4}^{+0.5}\times 10^{-7}$ \\
$1.19_{-0.03}^{+0.09}$&\textit{Swift}/XRT & $5.5\pm 1.4\times 10^{-13}$ &$5.7\pm 1.4\times 10^{-8}$ \\
$1.71_{-0.17}^{+0.76}$&\textit{Swift}/XRT & $2.8_{-0.8}^{+0.9}\times 10^{-13}$ &$2.9_{-0.8}^{+0.9}\times 10^{-8}$ \\
$4.71_{-0.06}^{+0.35}$&\textit{Swift}/XRT & $1.1_{-0.4}^{+0.5}\times 10^{-14}$ &$1.2_{-0.4}^{+0.5}\times 10^{-8}$ \\
$13.50_{-0.24}^{+0.33}$&\textit{XMM-Newton} & $1.14\pm 0.17 \times 10^{-14}$ &$1.18_{-0.18}^{+0.17} \times 10^{-9}$\\
$25.3_{-1.5}^{+0.6}$&\textit{Chandra} &$2.6\pm 1.3 \times 10^{-15}$& $3.4\pm 1.7\times 10^{-10}$\\
$37.01_{-0.59}^{+0.64}$&\textit{XMM-Newton} & $1.3 _{-1.0}^{+0.6}  \times 10^{-15}$ &$1.4 _{-1.0} ^{+0.6} \times 10^{-10}$\\
\bottomrule
\end{tabular}
\begin{tablenotes}
\footnotesize
\item [a] Flux in the energy band 0.3--10 keV. 
\item [b] Flux density at 1 keV. 
\end{tablenotes}
\end{threeparttable}
\end{table}

\clearpage
\begin{ThreePartTable}
\centering
\small
\begin{longtable}{c c c c }
\caption{ \textbf{Photometric observations of GRB 230307A.} Magnitudes are reported in the AB system and are not corrected for Galactic extinction. Errors represent the $1\sigma$ uncertainties. Upper limits are given at $3\sigma$ level.
}\label{tab:photometry - UVOIR}\\
\toprule
$\Delta t$ (day)& Telescope&Filter& Magnitude\\%
\midrule
0.02&\textit{TESS}&\textit{Red}&${18.15}\pm{0.07}$\\%
0.04&\textit{TESS}&\textit{Red}&${17.82}\pm{0.06}$\\%
0.07&\textit{TESS}&\textit{Red}&${17.79}\pm{0.06}$\\%
0.09&\textit{TESS}&\textit{Red}&${17.82}\pm{0.06}$\\%
0.11&\textit{TESS}&\textit{Red}&${17.63}\pm{0.06}$\\%
0.13&\textit{TESS}&\textit{Red}&${18.21}\pm{0.07}$\\%
0.16&\textit{TESS}&\textit{Red}&${18.12}\pm{0.07}$\\%
0.19&\textit{TESS}&\textit{Red}&${18.28}\pm{0.08}$\\%
0.23&\textit{TESS}&\textit{Red}&${18.29}\pm{0.07}$\\%
0.26&\textit{TESS}&\textit{Red}&${18.45}\pm{0.08}$\\%
0.43&RASA36&\textit{r}&${19.4}\pm{0.6}$\\%
0.75&KMTNet/SSO&\textit{R}&${20.78}\pm{0.05}$\\%
0.97&UVOT&\textit{u}&${22.2}\pm{0.9}$\\%
1.12&KMTNet/SAAO&\textit{R}&${20.90}\pm{0.05}$\\%
1.20&PRIME&\textit{H}&${20.21}\pm{0.15}$\\%
1.20&PRIME&\textit{J}&${20.74}\pm{0.11}$\\%
1.20&PRIME&\textit{Y}&${20.60}\pm{0.14}$\\%
1.20&PRIME&\textit{Z}&$>{20.5}$\\%
1.20&UVOT&\textit{white}&${22.1}\pm{0.2}$\\%
1.60&UVOT&\textit{white}&${23.0}\pm{0.3}$\\%
1.82&KMTNet/SSO&\textit{I}&${21.20}\pm{0.10}$\\%
1.82&KMTNet/SSO&\textit{R}&${21.42}\pm{0.05}$\\%
2.13&KMTNet/SAAO&\textit{R}&${21.79}\pm{0.05}$\\%
2.14&KMTNet/SAAO&\textit{I}&${21.70}\pm{0.10}$\\%
2.35&Gemini&\textit{r}&${22.0}\pm{0.3}$\\%
2.36&SOAR&\textit{z'}&${21.50}\pm{0.12}$\\%
2.38&KMTNet/CTIO&\textit{R}&${22.46}\pm{0.06}$\\%
2.39&KMTNet/CTIO&\textit{I}&${22.0}\pm{0.2}$\\%
2.77&KMTNet/SSO&\textit{R}&${22.46}\pm{0.06}$\\%
3.12&KMTNet/SAAO&\textit{R}&${22.61}\pm{0.06}$\\%
3.36&SOAR&\textit{z'}&${22.6}\pm{0.2}$\\%
4.12&KMTNet/SAAO&\textit{R}&${23.30}\pm{0.07}$\\%
4.76&KMTNet/SSO&\textit{R}&${23.65}\pm{0.08}$\\%
5.12&KMTNet/SAAO&\textit{R}&${23.61}\pm{0.08}$\\%
6.13&KMTNet/SAAO&\textit{R}&${23.67}\pm{0.08}$\\%
7.40&Gemini&\textit{r}&${24.90}\pm{0.10}$\\%
7.40&XSH&\textit{K}&${22.0}\pm{0.6}$\\%
8.34&Gemini&\textit{z}&${24.3}\pm{0.2}$\\%
15.33&Gemini&\textit{J}&$>{23.5}$\\%
15.41&Gemini&\textit{r}&$>{25}$\\%
17.33&Gemini&\textit{J}&$>{23.0}$\\%
28.9&\textit{JWST}&\textit{F444W}&${24.52}\pm{0.01}$\\%
28.9&\textit{JWST}&\textit{F356W}&${25.26}\pm{0.02}$\\%
28.9&\textit{JWST}&\textit{F277W}&${26.12}\pm{0.02}$\\%
28.9&\textit{JWST}&\textit{F150W}&${27.9}\pm{0.3}$\\%
28.9&\textit{JWST}&\textit{F115W}&${28.2}\pm{0.3}$\\%
28.9&\textit{JWST}&\textit{F070W}&${28.7}\pm{0.3}$\\%
29.4&\textit{HST}&\textit{F140W}&$>{27.2}$\\%
29.4&\textit{HST}&\textit{F105W}&$>{27.2}$\\%
55.8&\textit{HST}&\textit{F140W}&$>{27.2}$\\%
55.8&\textit{HST}&\textit{F105W}&$>{27.2}$\\%
61.4&\textit{JWST}&\textit{F444W}&${26.94}\pm{0.08}$\\%
61.4&\textit{JWST}&\textit{F277W}&${28.6}\pm{0.3}$\\%
61.4&\textit{JWST}&\textit{F150W}&$>{28.9}$\\%
61.4&\textit{JWST}&\textit{F115W}&$>{28.7}$\\%
\bottomrule
\end{longtable}
\end{ThreePartTable}

\begin{table}
\caption{\textbf{Radio observations from ATCA.} Errors represent the $1\sigma$ uncertainties. Upper limits are given at $3\sigma$ level. 
}
\label{tab:radio}
\centering
\begin{threeparttable} 
\begin{tabular}{ccc}
\toprule
$\Delta t$&Frequency&Flux density\\%
(day)&(GHz)&($\mu$Jy)\\%
\midrule
4.5&5.5&${162}\pm{24}$\\%
4.5&9.0&${130}\pm{27}$\\%
10.8&5.5&$<{100}$\\%
10.8&9.0&$<{52}$\\%
10.8&16.7&$<{48}$\\%
10.8&21.2&$<{104}$\\%
25.7&5.5&$<{50}$\\%
25.7&9.0&$<{40}$\\%
\bottomrule
\end{tabular}
\end{threeparttable} 
\end{table}
\clearpage

\section*{References}
\vspace{4em}

\bibliographystyle{naturemag}
\bibliography{main}

\end{document}